\documentclass[10pt,a4paper]{article}
\usepackage[utf8]{inputenc}
\usepackage[english]{babel}
\usepackage{graphpap,amsmath,amsfonts,amssymb,epsfig,makeidx,graphicx,textcomp,cite,citeall}
\usepackage[left=1.5cm, right=1.5cm, top=1.5cm, bottom=1.5cm]{geometry}

\begin{document}
	\textbf{\Large{\begin{center}{Compensation and its systematics in spin-1/2 Ising trilayered triangular ferrimagnet}\end{center}}}
	
	\begin{center}
		{\normalsize Soham Chandra}\footnote{E-mail: soham.rs@presiuniv.ac.in; sohamc07@gmail.com}\\
		{\it Department of Physics, Presidency University}\\
		{\it 86/1 College Street, Kolkata-700073, India}
	\end{center}
\vspace{20pt}
\begin{abstract}
Trilayered, Ising, spin-$1/2$, ferrimagnets are an interesting subject for simulational studies for they show compensation effect. A Monte Carlo study on such a system with sublayers on triangular lattice is performed in the current work. Three layers, making up the bulk, is formed completely by either A or B type of atoms. The interactions between like atoms (A-A; B-B) are ferromagnetic and between unlike ones (A-B) are anti-ferromagnetic. Thus the system has three coupling constants and manifests into two distinct trilayer compositions: AAB and ABA. Metropolis single spin flip algorithm is employed for the simulation and the location of the critical points (sublattice magnetisations vanish, leading to zero bulk magnetisation) and the compensation points (bulk magnetisation vanishes but nonzero sublattice magnetisations exist) are estimated. Close range simulations with variable lattice sizes for compensation point and Binder's cumulant crossing technique for critical points are employed for analysis and conditions for the existence of compensation points are determined. Comprehensive phase diagrams are obtained in the Hamiltonian parameter space and morphological studies at critical and compensation temperatures for both the configurations are also reported. The alternative description in terms of \textit{Inverse absolute of reduced residual magnetisation} and \textit{Temperature interval between Critical and Compensation temperatures} is also proposed and compared with traditional simulational results. Such simulational studies and the proposed systematics of compensation effect are useful in designing materials for specific technological applications.
\end{abstract}
\vspace{50pt}
\newpage
\begin{center}{\bf \Large I. Introduction}\end{center}

Ferrimagnetism was discovered in 1948 \cite{Cullity} and studies on ferrimagnets have revealed unique properties and phase diagrams \cite{Connell,Camley}. Each of the substructures of a layered ferrimagnet, may have different thermal dependencies for magnetization and such different behaviors, combined, leads to interesting phenomena such as \textit{compensation}, i.e., temperature(s) below the critical point for which total magnetization of the bulk becomes zero while substructures retain their magnetic order. Compensation is not related to criticalilty but some physical properties like the magnetic coercivity exhibits singularity at the compensation point \cite{Connell,Ostorero}. For thermomagnetic recording devices, ferrimagnetic materials which have their Compensation points around room temperature and also have strong temperature dependence of coercive field around the compensation point, are good candidates \cite{Connell}. Some ferrimagnets even have their compensation points near room temperature \cite{Ostorero}, making them ideal for magneto-optical drives.
Layered ferrimagnetic materials owing to their enhanced surface-to-volume ratio, present unique features, quite different from the bulk. Now-a-days, with atomic layer deposition
(ALD) \cite{George}, pulsed laser deposition (PLD) \cite{Singh2}, metalorganic chemical vapor deposition (MOCVD) \cite{Stringfellow} and molecular-beam epitaxy (MBE) \cite{Herman}, experimental growth of bilayered \cite{Stier}, trilayered \cite{Leiner}, and multilayered \cite{Sankowski,Maitra} systems with desired characteristics has been achieved. The main interest of this article, is to find out the dependence of compensation phenomenon on interaction strengths of the Hamiltonian of a trilayered, spin 1/2, Ising \cite{Ising}  ferrimagnet with sublayers having triangular Bravais lattice structure.\\ 
\indent The challenge for experimental studies of model magnetic interactions in ideal systems is two-fold: (a) finding such naturally ocurring materials and; (b) in absence of (a), artificially engineering them. Recently Atomic physicists have experimentally realized artificial gauge potentials for bulk \cite{Schweikhard,Lin1,Lin2} and optical lattice systems \cite{Aidelsburger,Struck,Jimenez}. Such advancements have provided a solution to this problem by engineering relevant spin interactions in quantum simulators \cite{Butula} using e.g. ultracold bosonic quantum gases in optical lattices \cite{Lewenstein}. In such versatile experimental set-ups: precise quantum spin-state control, engineered spin-spin coupling and deterministic spin localization permits us to control certain parameters like lattice spacing and geometry, and spin-spin interaction strength and range. The work in reference \cite{Struck} is performed on triangular lattice. In \cite{Britton}, such methodology is successfully implemented by a Penning trap apparatus where laser-cooled ${}^{9}Be^{+}$ ions ($\sim300$ spins) naturally form a stable 2D Coulomb crystal on a \textit{triangular} lattice. In such a lattice, each ion is a spin-$1/2$ system (qubit) over which the authors used high-fidelity quantum control \cite{Biercuk} and a spin-dependent optical dipole force (ODF) to engineer a continuously tunable Ising-type spin-spin coupling. The strong geometrical anisotropy of van der Waals layered crystals leads to a significant difference in magnitude between intralayer and interlayer exchange couplings. An example is $CrI_{3}$, in which, magnetization and susceptibility measurements show \cite{McGuire1} that bulk $CrI_{3}$ is a strongly anisotropic ferromagnet below the Curie temperature $(=61\text{ K})$, with saturation magnetization consistent with $s=3/2$ state of the $Cr$ atoms. But when the bulk is reduced to a few layers, we observe intra-layer ferromagnetic and inter-layer antiferromagnetic interactions. The sublattice magnetizations point perpendicular to the plane \cite{McGuire2}.

Using \textit{Ising interactions}, thin films have been studied in literature by computatinal and analytical techniques e.g. by equilibrium Monte Carlo (MC) simulations in \cite{Laosiritaworn,Albano}, by mean-field theory (MFT) in \cite{Lubensky}, by effective-
field theory (EFT) in \cite{Kaneyoshi}, by series-expansion
method in \cite{Oitmaa}, by renormalization-group (RG) method in \cite{Ohno}, by spin-fluctuation theory in \cite{Benneman}, by exact recursion equation on the Bethe lattice in \cite{Albayrak} and by pair approximation method (PAM) in \cite{Balcerzak,Szalowski2}. 
But methods for exact solutions of spin systems are very few. That is why numerical and approximate investigations, in Compensation studies are quite significant. Recent investigations on compensation on Ising spin-$1/2$ trilayered ferrimagnets include: In \cite{Diaz1}, by MFA and EFA and in \cite{Diaz3}, by MC simulations with Wolff single cluster Algorithm, a spin-1/2 pure Ising trilayer on \textit{square} lattice was investigated and the authors have shown that under certain range of different types of interaction strengths, different temperature dependencies of sublattice magnetisations cause the compensation point to appear. In \cite{Albayrak}, investigations on trilayer Bethe lattice with same spin, $(1/2, 1/2, 1/2)$ and mixed spins $(1/2, 1, 1/2); (1, 1/2, 1); (1/2, 3/2, 1/2)$ are performed by recursion relations. These studies \cite{Albayrak} show availability of a number of phases namely ferromagnetic, antiferromagnetic, surface ferromagnetic, \textit{compensated}, mixed phase and surface antiferromagnetic. In \cite{Chandra}, two properties of the bulk for an Ising spin-$1/2$ square trilayered ferrimagnet are introduced. One is Inverse absolute of reduced residual magnetisation (IARRM), and the other is temperature interval between Critical and Compensation temperatures (TICCT). There, for those two unconventional quantities, possible mathematical forms of dependences on relative interaction strengths in the Hamiltonian, are proposed.\\
\indent But current simulational studies on trilayered ferrimagnets are mostly centered around square sublattices. Change in the underlying lattice structure may be significant since characterictics of any crystalline material depend on its lattice symmetry e.g. critical temperature of a magnetic system changes with a change in the coordination number. Especially in light of the recent experimental realizations \cite{Schweikhard,Lin1,Lin2,Aidelsburger,Struck,Jimenez,Butula,Lewenstein,Britton,Biercuk,McGuire1,McGuire2}, it would be interesting to study the behaviour of the critical ($T_{crit}$) and compensation temperatures ($T_{comp}$) on controlling parameters of the system and the resulting phase diagram of a spin-$1/2$ Ising trilayered system on \textit{triangular} lattice. Simulational studies on such systems help technologists in choosing and designing efficient materials for specific purposes e.g. magnetic refrigeration by magnetocaloric effect. The alternative description, in terms of IARRM and TICCT, for such systems is also obtained in the current study.
\\ \indent The rest of the article is arranged as follows. In Sec. II, the model of this study is described. In Sec. III, the details of the MC simulation scheme are provided. In Sec. IV, the simulational results for AAB and ABA configurations are discussed. Finally, in Sec. V, the conclusion of this study is provided.
\noindent \begin{center}{\bf \Large II. Model}\end{center}

\begin{figure}[!htb]
	\begin{center}
		\begin{tabular}{c}
			(a)
			\resizebox{8cm}{!}{\includegraphics[angle=0]{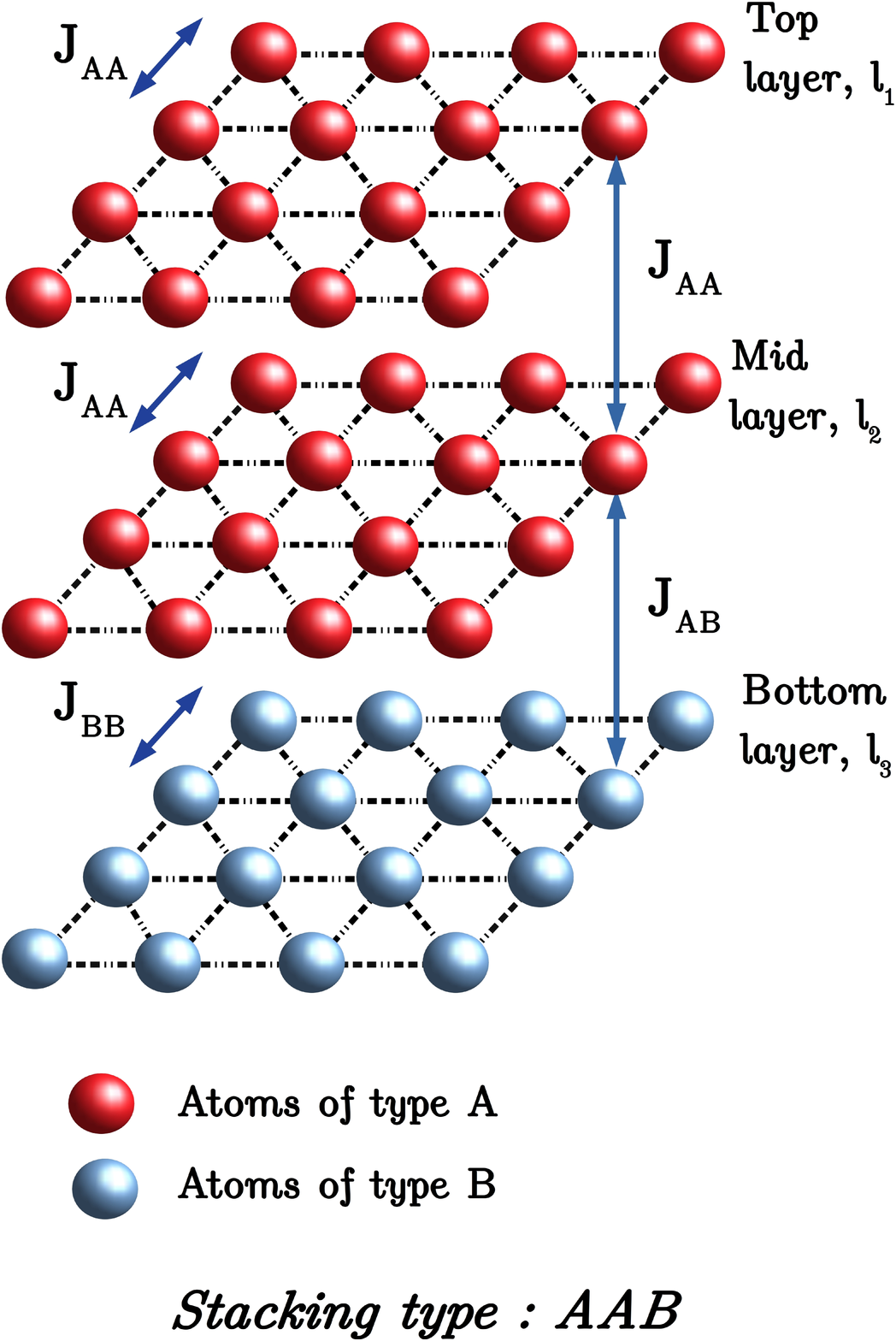}}
			(b)
			\resizebox{8cm}{!}{\includegraphics[angle=0]{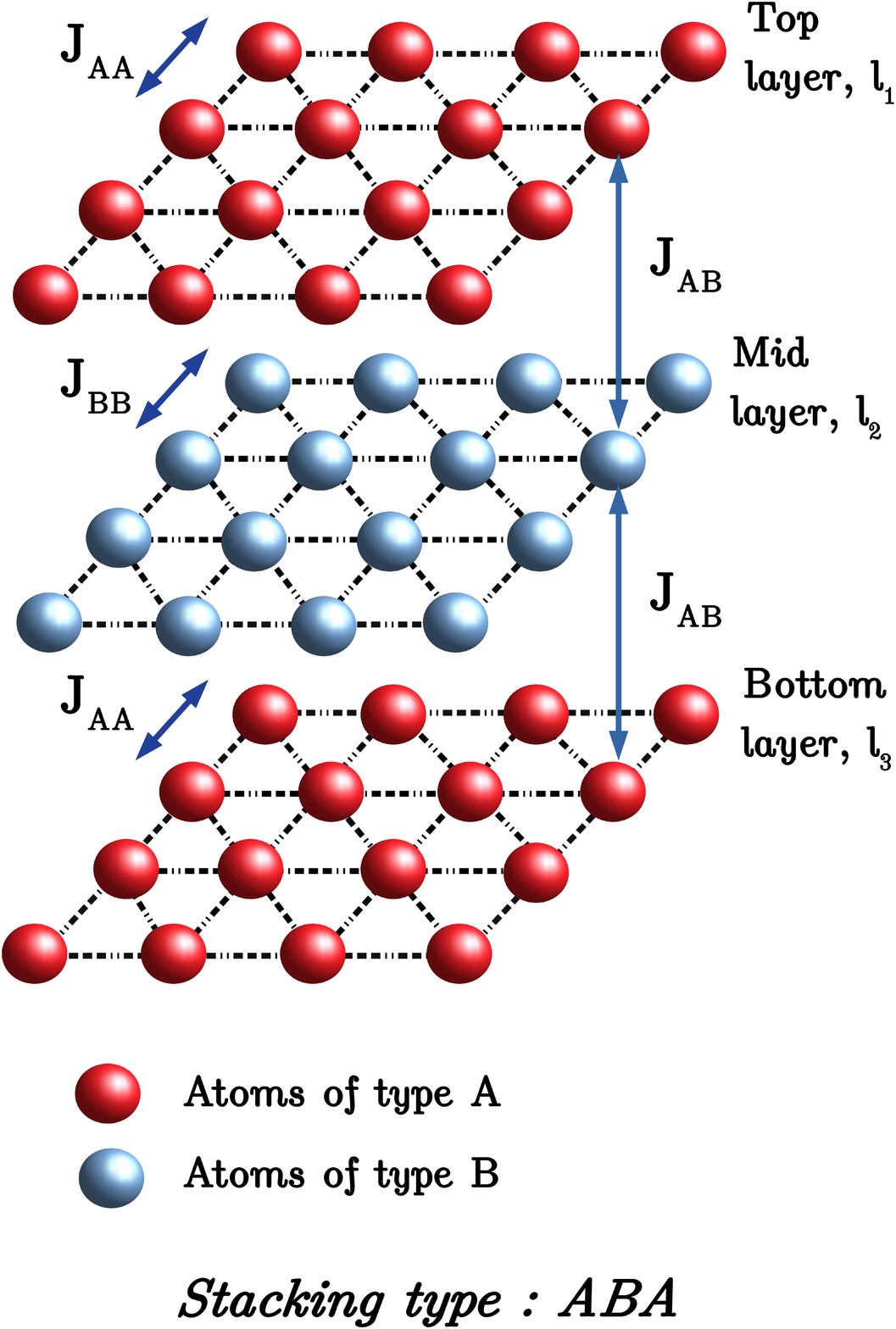}}
		\end{tabular}
		\caption{(Colour Online) Two distinct trilayer configurations: (a) AAB (with $J_{11}=J_{22}=J_{12}=J_{AA}$, $J_{33}=J_{BB}$ and $J_{23}=J_{AB}$)(b) ABA (with $J_{11}=J_{33}=J_{AA}$, $J_{22}=J_{BB}$ and $J_{12}=J_{23}=J_{AB}$).}
		\label{fig_lattice_structure}
	\end{center}
\end{figure}

The \textit{layered} Ising superlattice of this study, contains three magnetic layers. These three sublayers have triangular Bravais lattice symmetry. Each of these layers is composed completely by either, A or B, one of the two types of theoretical atoms. The magnetic interactions between nearest neighbours are Ising-like and their natures are:\\
(a) A-A $\to$ Ferromagnetic\\
(b) B-B $\to$ Ferromagnetic\\
(c) A-B $\to$ Anti-ferromagnetic,\\
which results in Two different configurations: (i) AAB [Figure \ref{fig_lattice_structure}(a)] and (ii) ABA [Figure \ref{fig_lattice_structure}(b)].

The Hamiltonian for such a trilayered ferrimagnetic system, using nearest neighbour Ising mechanics \cite{Ising}, is (with all the $S^{z}$'s are $z$ components of spin moments on lattice sites):
\begin{equation}
\label{eq_Hamiltonian}
H=-J_{11}\sum_{<t,{t}^{\prime }>}S_{t}^{z}S_{{t}^{\prime }}^{z}-J_{22}\sum_{<m,{m}^{\prime }>}S_{m}^{z}S_{{m}^{\prime }}^{z}-J_{33}\sum_{<b,{b}^{\prime }>}S_{b}^{z}S_{{b}^{\prime }}^{z}-J_{12}\sum_{<t,m>}S_{t}^{z}S_{m}^{z}-J_{23}\sum_{<m,b>}S_{m}^{z}S_{b}^{z}
\end{equation}
where $J_{11}$ is the coupling strength between nearest neighbour sites on top layer, $l_{1}$ and similarly introduced are $J_{22}$ for the mid-layer, $l_{2}$ and $J_{33}$ for the bottom-layer, $l_{3}$. $J_{12}$ and $J_{23}$ are the inter-layer nearest neighbour coupling strengths between the top-mid and mid-bottom layers. Summation indices $(t,{t}^{\prime})$; $(m,{m}^{\prime})$ and $(b,{b}^{\prime})$ are respectively for the lattice sites on the top-layer, $l_{1}$; mid layer, $l_{2}$ and bottom-layer, $l_{3}$ and $\langle t,{t}^{\prime }\rangle$, $\langle m,{m}^{\prime }\rangle$, $\langle b,{b}^{\prime }\rangle$ denote summations over all nearest-neighbor pairs in the same layer and $\langle t,m\rangle$, $\langle m,b\rangle$ are summations over nearest-neighbor pairs in vertically adjacent layers. In Equation (\ref{eq_Hamiltonian}), the first, second and third terms respectively are for the intra-planar ferromagnetic contributions from the top, mid and bottom layers. The fourth and the fifth terms originate due to the nearest neighbour inter-planar antiferromagnetic interactions, between top \& mid and mid \& bottom layers.\\
\indent For the AAB type system, in Equation (\ref{eq_Hamiltonian}): $J_{11}>0$ , $J_{22}>0$, $J_{33}>0$ and $J_{12}>0$, $J_{23}<0$ and in terms of unique interactions: $J_{11}=J_{22}=J_{12}=J_{AA}$, $J_{33}=J_{BB}$ and $J_{23}=J_{AB}$. And for the ABA variant, the nature of the coupling strengths in Equation (\ref{eq_Hamiltonian}) are: $J_{11}>0$ , $J_{22}>0$, $J_{33}>0$ and $J_{12}<0$, $J_{23}<0$ and $J_{11}=J_{33}=J_{AA}$, $J_{22}=J_{BB}$ and $J_{12}=J_{23}=J_{AB}$. There is no out-of-plane interaction term between the top and bottom layers in the Hamiltonian, that is why periodic boundary conditions in-plane and open boundary conditions along the vertical are considered.
\vspace{20pt}
\noindent \begin{center}{\bf \Large III. Simulation scheme}\end{center}

\begin{figure}[!htb]
	\begin{center}
		\begin{tabular}{c}
			\resizebox{10cm}{!}{\includegraphics[angle=0]{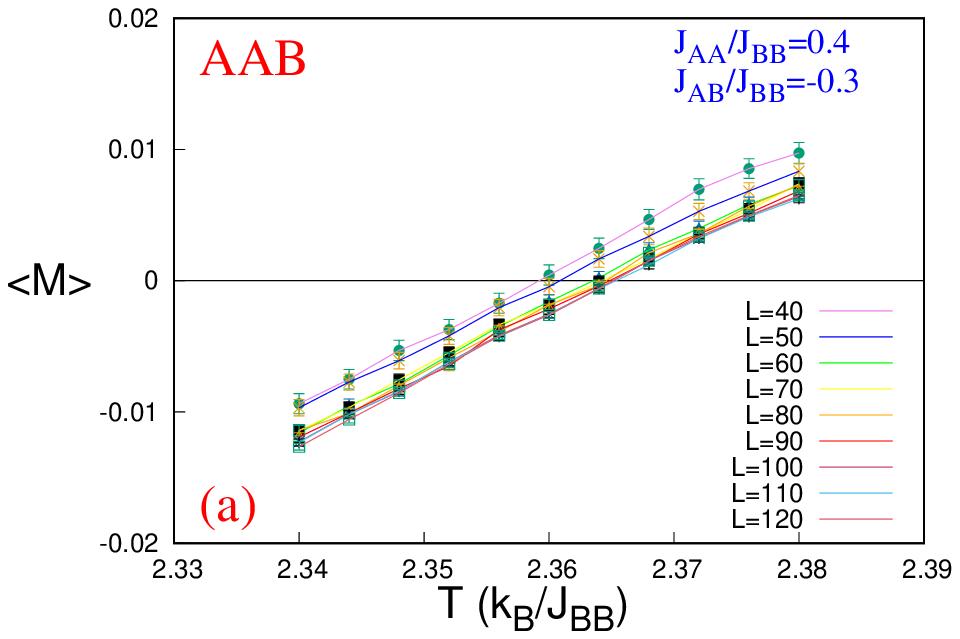}}
			\resizebox{10cm}{!}{\includegraphics[angle=0]{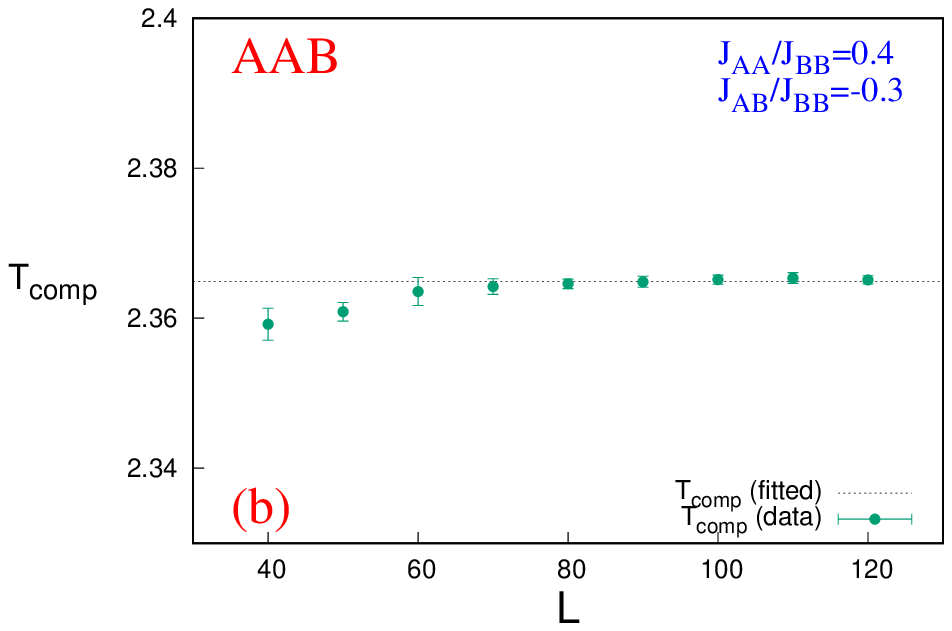}}\\
			\resizebox{10cm}{!}{\includegraphics[angle=0]{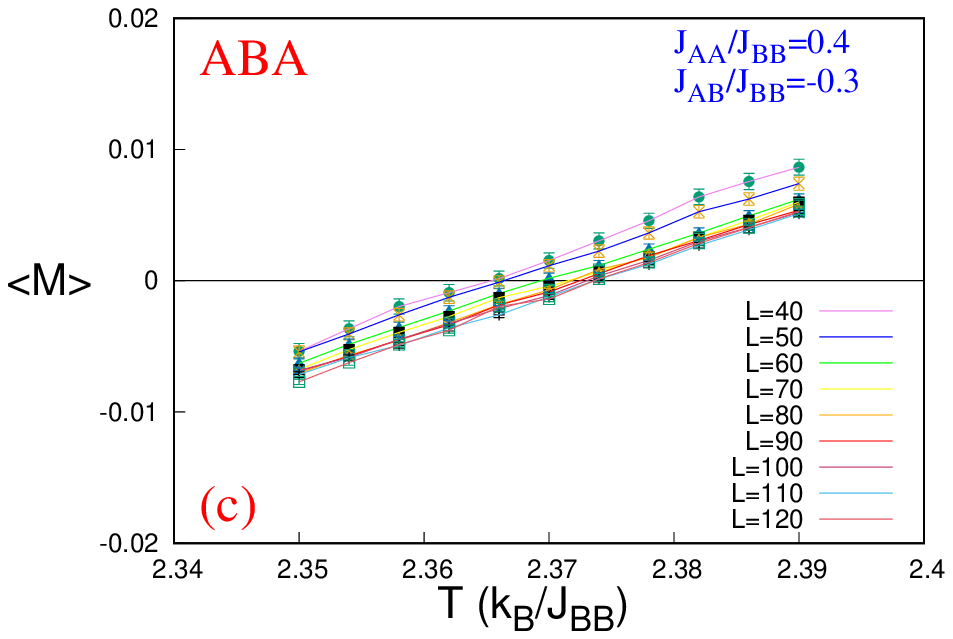}}
			\resizebox{10cm}{!}{\includegraphics[angle=0]{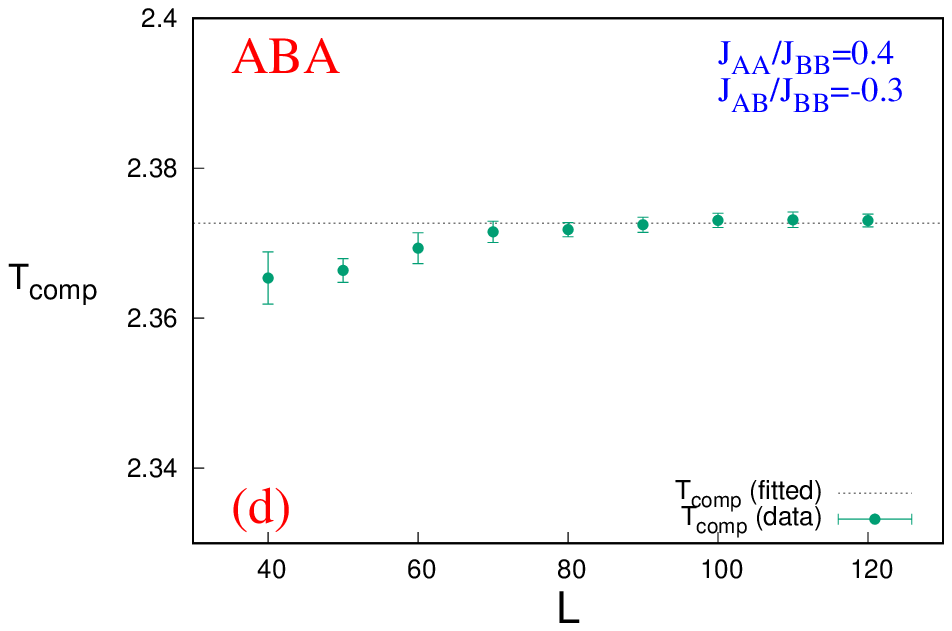}}
		\end{tabular}
		\caption{ (Colour Online) Simulations in the vicinity of $T_{comp}$, for (a) AAB (c) ABA configurations and fitting of data for (b) AAB (d) ABA configuartions ($J_{AA}/J_{BB}=0.4$ and $J_{AB}/J_{BB}=-0.3$ for both). The dimensionless compeansation temperatures, came out to be $2.3649\pm0.0011$ for AAB and $2.3727\pm0.0015$ for ABA configuration, which are marked by dashed lines in (b) and (d)}		
		\label{fig_comp_simfit}
	\end{center}
\end{figure}

The model, described in Section II, is simulated by employing the Monte Carlo simulations with Metropolis single spin-flip algorithm \cite{Binder-Landau}, with each plane having $L^{2}$ sites where $L=100$. For $L\geqslant70$ [Refer to Figure \ref{fig_comp_simfit}], the compensation point remains confined within a narrow band, around a stable value. Thus the lattice size considered in this study is sufficient for obtaining statistically reliable results. The simulation started from a high temperature paramagnetic phase, having randomly selected 50\% spin projections, $S_{i}^{z}=+1$ and the rest with $S_{i}^{z}=-1$ (Using $1$ instead of $1/2$ rescales the coupling constants). At a fixed temperature $T$, the Metropolis rate \cite{Metropolis,Newman}, of Equation [\ref{eq_metropolis}], governs the spin flipping from $S_{i}^{z}$ to $-S_{i}^{z}$:
\begin{equation}
\label{eq_metropolis}
P(S_{i}^{z} \to -S_{i}^{z}) = \text{min} \{1, \exp (-\Delta E/k_{B}T)\}
\end{equation}
where $\Delta E$ is the associated change in internal energy in flipping the $i$-th spin projection from $S_{i}^{z}$ to $-S_{i}^{z}$ with Boltzmann constant, $k_{B}$ set to $1$. Such $3L^{2}$ random single-spin updates make up one Monte Carlo Sweep (MCS) for the entire system. This \textit{one MCS} is the unit of time in the current study. At every temperature step, the first $5\times10^{4}$ MCS (that is equivalent to allowance of a long enough \textit{time}) were discarded for thermalization and then $N=100$ \textit{uncorrelated} microstates for thermal averages were considered after accounting for integrated autocorrelation time \cite{Newman}. The temperatures of the systems are measured in units of $J_{BB}/k_{B}$. The choice of total MCS (thus the number of uncorrelated states) is made depending upon the available computational resources and statistical reliability.\\
\indent Both, ABA and AAB configurations, were observed for ten values of $J_{AA}/J_{BB}$, starting from $0.1$ to $1.0$ with an interval of $0.1$ and for each fixed value of $J_{AA}/J_{BB}$ , $J_{AB}/J_{BB}$ was decreased from $-0.1$ to $-1.0$ with an interval of $-0.1$. For \textit{each combination} of $J_{AA}/J_{BB}$ and $J_{AB}/J_{BB}$ , the time (or, ensemble) averages of the following quantities were calculated at each of the temperature points, in the following manner:\\
\textbf{(1) Sublattice magnetisations} for top, mid and bottom layers calculated, identically, at $i$-th uncorrelated configuration after equilibration, denoted by $M_{qi}$, by:
\begin{equation}
M_{qi}=\frac{1}{L^{2}}\sum_{x,y=1}^{L} \left( S_{qi}^{z}\right)_{xy}
\end{equation}
and the sum extends over all sites in each of the planes as $x$ and $y$ denote the co-ordinates of a spin on $q$-th sublayer and runs from $1$ to $L$ (which is $100$, in this study). Then the time (or, ensemble) average, from the $N$ uncorrelated configurations is obtained as follows:
\begin{equation}
\langle M_{q}\rangle=\dfrac{1}{N}\sum_{i=1}^{N}M_{qi}
\end{equation} 
where $q$ is to be replaced by $t,m\text{ or }b$ for top, mid and bottom layers and  $\langle\cdots\rangle$ denotes a time average (equivalently ensemble average) after attaining equilibrium.\\
\textbf{(2)} \textbf{Average magnetisation of the trilayer} by $\langle M\rangle=\dfrac{1}{3}\left(\langle M_{t}\rangle+\langle M_{m}\rangle+\langle M_{b}\rangle\right)$\\
\textbf{(3)} After attaining equilibrium, the \textbf{fluctuation in magnetisation, $\Delta M$} is calculated from the uncorelated $N$ MCS by:
\begin{equation}
{\Delta M}=\sqrt{\dfrac{1}{N-1} \sum_{i=1}^{N} (M_{i}-\overline{M})^{2} }
\end{equation}
where $M_{i}$ is the value of magnetisation of the whole system, calculated from the $i$-th uncorrelated MCS and $\overline{M}$ is the average value of total magnetisation calculated over the total $N$ uncorrelated MCS after equilibration. The errors associated
with the magnetizations and fluctuation in magnetization are estimated by Jackknife method \cite{Newman}.\\
\indent After the temperature sequence of the above mentioned quantities is stored for any specific combination of Coupling ratios, we can compute the values of Inverse Absolute of Reduced Residual Magnetisation (IARRM) and Temperature interval between Critical and Compensation temperatures (TICCT) as a function of the coupling strengths. The procedure follows.\\ 
\textbf{(4)} \textbf{Inverse Absolute of Reduced Residual Magnetisation (IARRM)} is determined at each combination of coupling strengths by calculating the absolute value of the ratio between intermediate maximum/minimum value of magnetization between critical and compensation temperature and the value of magnetization at the lowest simulational temperature ($\approx$ saturation magnetization).\\
\indent IARRM is thus a dimensionless quantity and denoting it by $\mu$, is defined by:
\begin{equation}
\mu\left(\dfrac{J_{AA}}{J_{BB}},\dfrac{J_{AB}}{J_{BB}}\right)=\left| \dfrac{M_{ext,int}}{M_{sat}}\right|
\end{equation}
\\
\textbf{(5)} \textbf{Temperature interval between Critical and Compensation temperatures (TICCT)} is calculated after determining the locations of Critical and Compensation temperatures on the temperature axis.\\
\indent TICCT, having the dimension of temperatre, is denoted by $\Delta T$, and is defined by:
\begin{equation}
\Delta T \left(\dfrac{J_{AA}}{J_{BB}},\dfrac{J_{AB}}{J_{BB}}\right)= T_{crit}-T_{comp}
\end{equation}
The analytical procedures for IARRM and TICCT are to be found, in detail, in Section IV(D).
\vspace{20pt}
\begin{center} {\Large \textbf {IV. Results}}\end{center}

\begin{figure}[!htb]
	\begin{center}
		\begin{tabular}{c}			
			\resizebox{10cm}{!}{\includegraphics[angle=0]{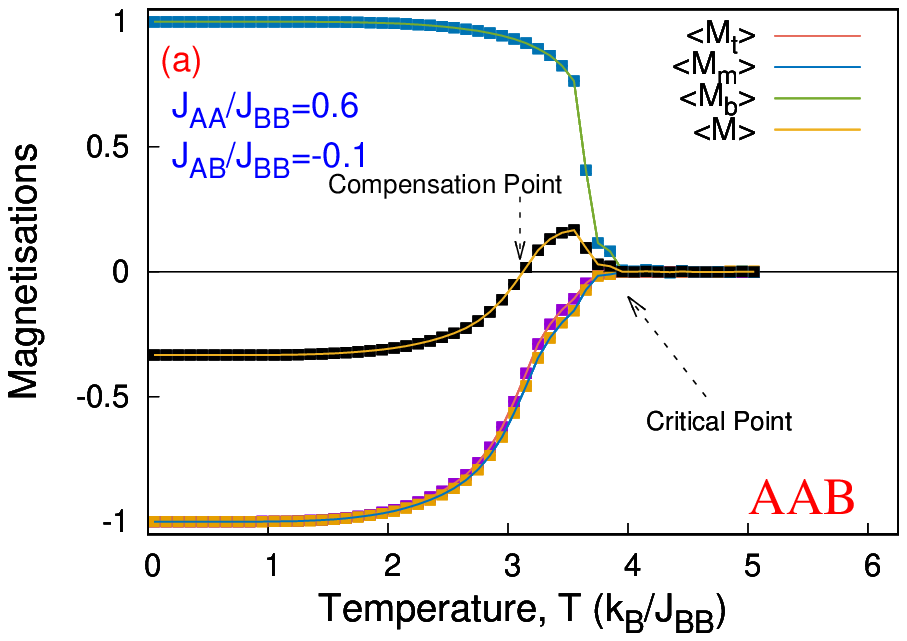}}
			\resizebox{10cm}{!}{\includegraphics[angle=0]{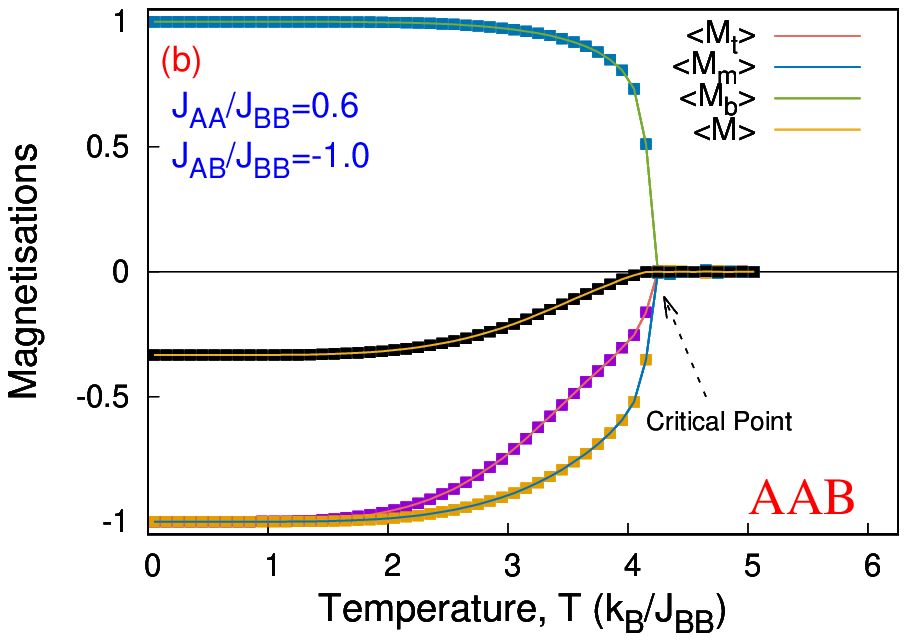}}
			\\
			\resizebox{10cm}{!}{\includegraphics[angle=0]{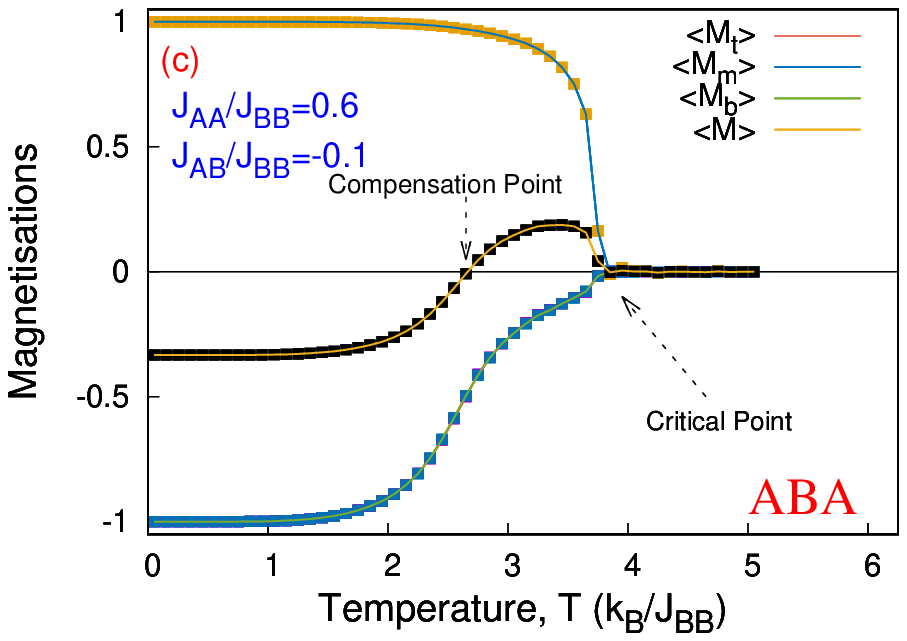}}
			\resizebox{10cm}{!}{\includegraphics[angle=0]{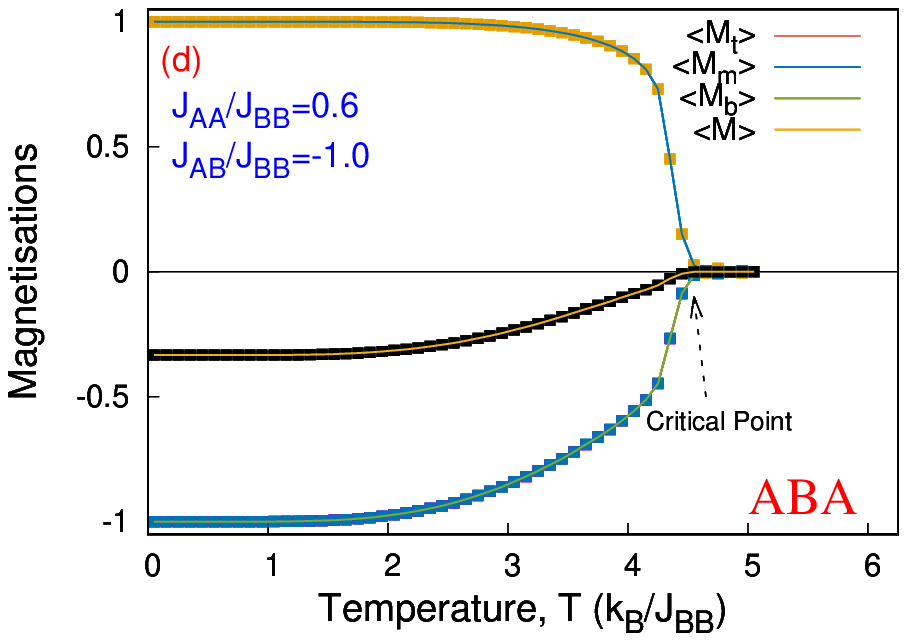}}
		\end{tabular}
		\caption{(Colour Online) Magnetisations as a function of dimensionless temperature: of AAB  configuration, (a) with compensation, (b) without compensation; and of ABA  configuration, (c) with compensation, (d) without compensation; with $J_{AA}/J_{BB}=0.6$ and $J_{AA}/J_{BB}=-0.1$ for those with compensation and $J_{AA}/J_{BB}=0.6$ and $J_{AA}/J_{BB}=-1.0$ for those without compensation. Where, the errorbars are not visible, they are smaller than the area of the point-markers. All these plots are obtained for a system of $3\times100\times100$ sites i.e. for $L=100$.}
		\label{fig_mag_response}
	\end{center}
\end{figure}

The thermodynamic and magnetic response of a trilayered triangular Ising ferrimagnet along with its morphology with MC single spin flip algorithm for both the distinct stackings are investigated. The effects of Hamiltonian parameters on the location and existence of compensation are observed and critical temperatures and finally obtained a phase diagram for both of them in the parameter space from MC data and from the mathematical relations of IARRM and TICCT.
\vspace{10pt}
\begin{center}{\Large \textbf {A. Magnetic response :}}\end{center}
\vspace{10pt}
In Figure \ref{fig_mag_response}, is shown the general trend of the behaviour of sublattice and average magnetizations of the system as a function of temperature for both type of configurations, AAB and ABA. $J_{AA}/J_{BB}=0.6$ and $J_{AA}/J_{BB}=-0.1$ are chosen for showing how they behave when compensation is present and $J_{AA}/J_{BB}=0.6$ and $J_{AA}/J_{BB}=-1.0$ for their behaviours in absense of compensation. All the figures are drawn for $L = 100$. 

\begin{figure}[!htb]
	\begin{center}
		\begin{tabular}{c}
			\resizebox{10cm}{!}{\includegraphics[angle=0]{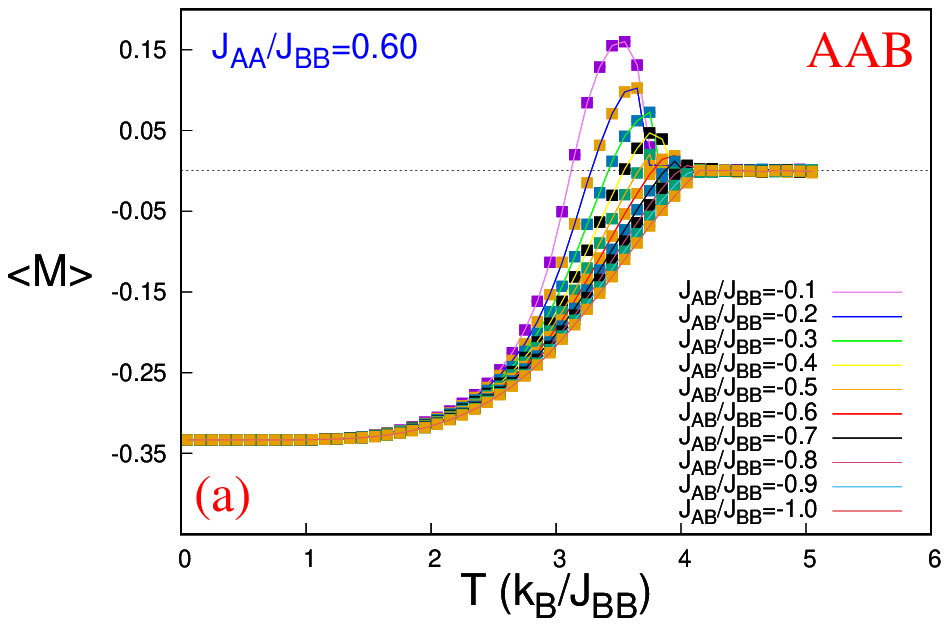}}
			\resizebox{10cm}{!}{\includegraphics[angle=0]{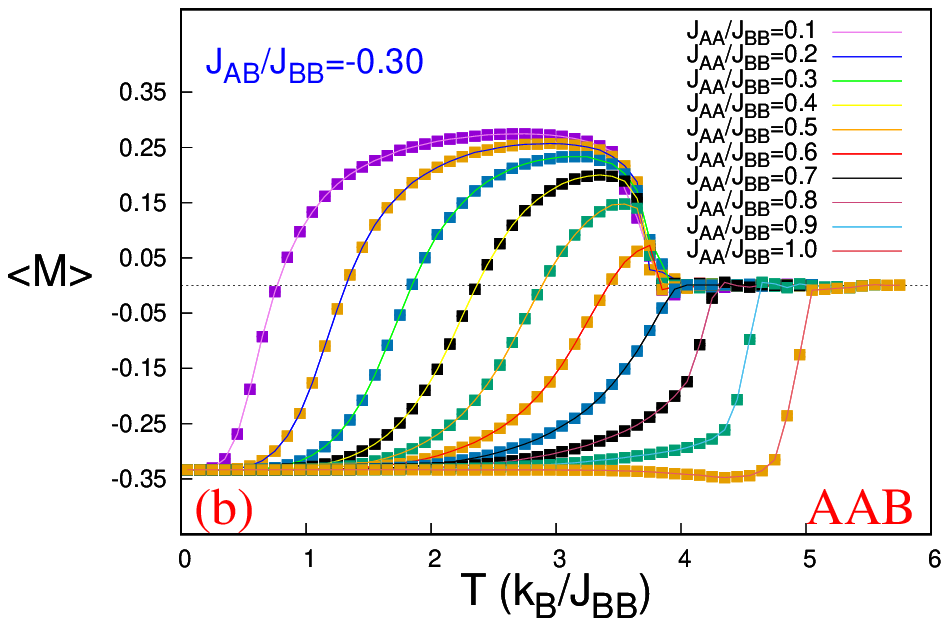}}
			\\
			\resizebox{10cm}{!}{\includegraphics[angle=0]{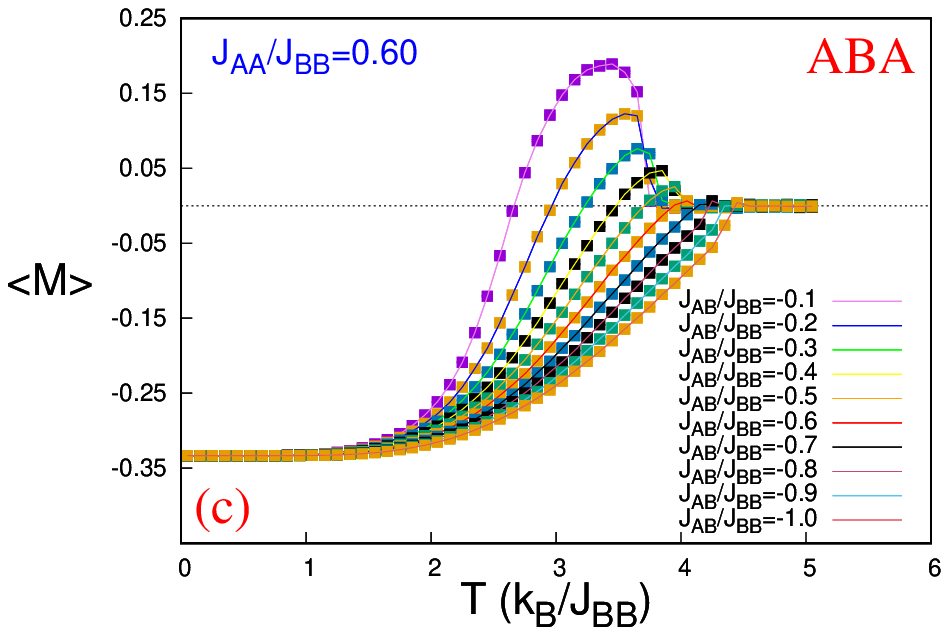}}	
			\resizebox{10cm}{!}{\includegraphics[angle=0]{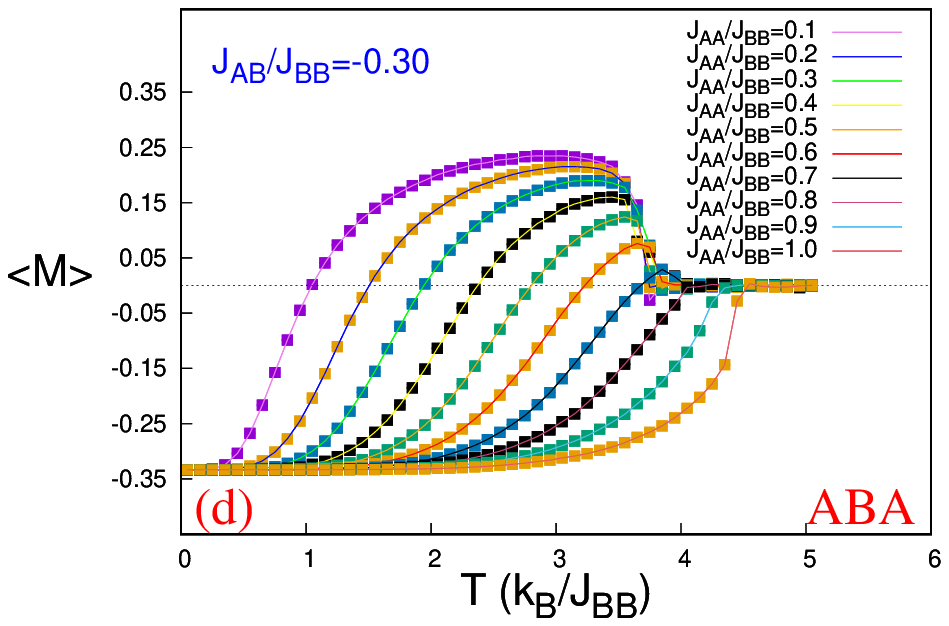}}
		\end{tabular}
		\caption{ (Colour Online) Average magnetisation of the bulk as a function of dimensionless temperature of: (a) AAB  configuration, with $J_{AA}/J_{BB}$ fixed at $0.6$ and variable $J_{AB}/J_{BB}$; (b) AAB  configuration, with $J_{AB}/J_{BB}$ fixed at $-0.3$ and variable $J_{AA}/J_{BB}$;; (c) ABA  configuration, with compensation; (d) ABA  configuration, without compensation; with $J_{AA}/J_{BB}=0.6$ and $J_{AA}/J_{BB}=-0.1$ for those with compensation and $J_{AA}/J_{BB}=0.6$ and $J_{AA}/J_{BB}=-1.0$ for those without compensation. Where, the errorbars are not visible, they are smaller than the area of the point-markers. All these plots are obtained for a system of $3\times100\times100$ sites i.e. for $L=100$.}
		\label{fig_mag_response_fixed}
	\end{center}
\end{figure}

Now $J_{AA}/J_{BB}$ is kept fixed to $0.6$ to see how compensation phenomenon changes under the variation of $J_{AB}/J_{BB}$ (varied from $-0.1$ to $-1.0$, decreased in steps of $-0.1$). On the same lines, $J_{AB}/J_{BB}$ is fixed to $-0.3$ and $J_{AA}/J_{BB}$ is varied (varied from $0.1$ to $1.0$, increased in steps of $0.1$), for both type of configurations. The results are shown in Figure \ref{fig_mag_response_fixed}.
\begin{figure}[!htb]
	\begin{center}
		\begin{tabular}{c}
			\resizebox{10cm}{!}{\includegraphics[angle=0]{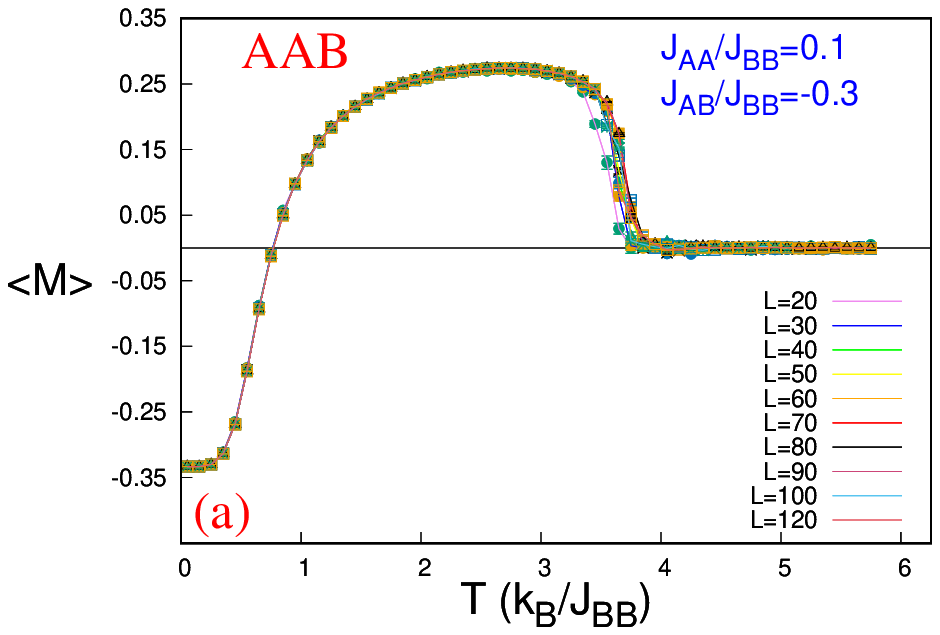}}
			\resizebox{10cm}{!}{\includegraphics[angle=0]{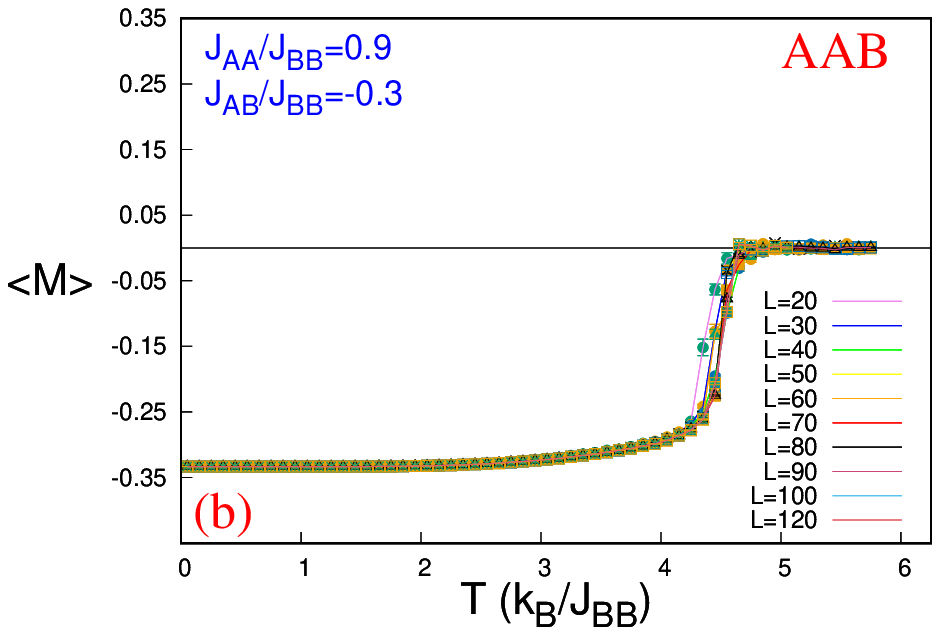}}
			\\
			\resizebox{10cm}{!}{\includegraphics[angle=0]{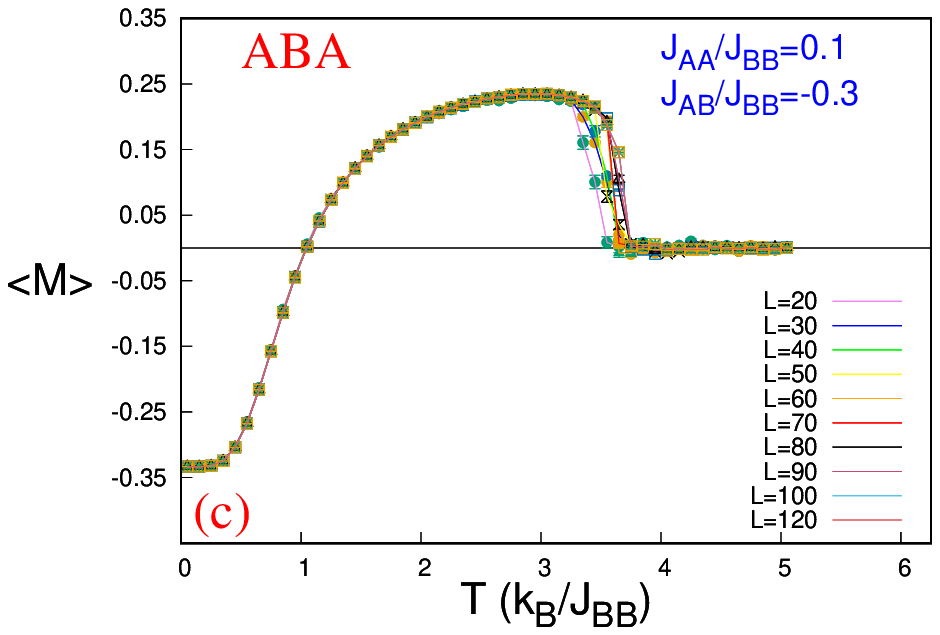}}
			\resizebox{10cm}{!}{\includegraphics[angle=0]{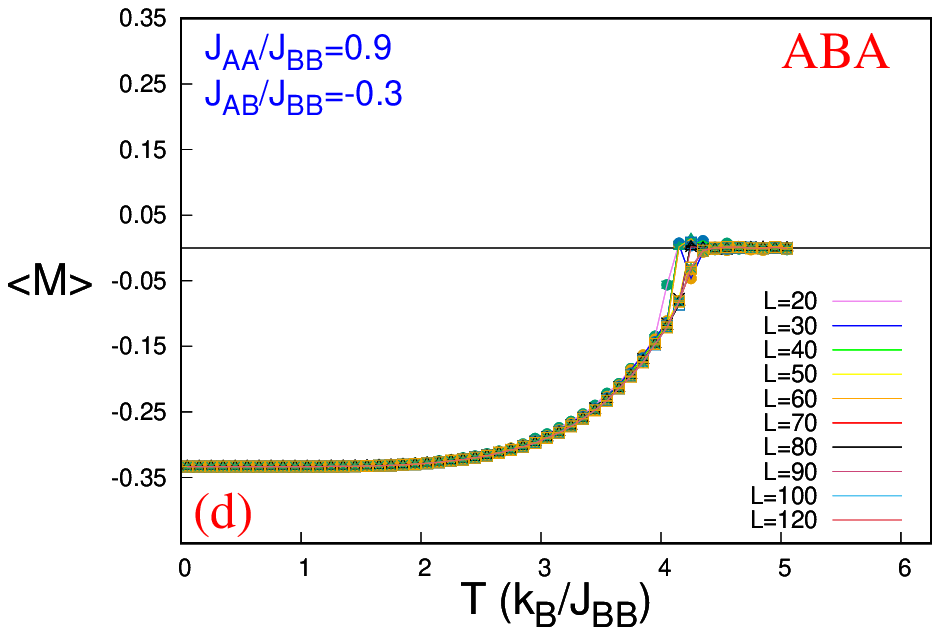}}
		\end{tabular}
		\caption{ (Colour Online) Average magnetisation of the bulk as a function of dimensionless temperature with variable lattice sizes, for: (a) AAB configuration with compensation, ; (b) AAB configuration, without compensation; (c) ABA configuration, with compensation; (d) ABA configuration, without compensation; with $J_{AA}/J_{BB}=0.1$ and $J_{AA}/J_{BB}=-0.3$ for those with compensation and $J_{AA}/J_{BB}=0.9$ and $J_{AA}/J_{BB}=-0.3$ for those without compensation. Where, the errorbars are not visible, they are smaller than the area of the point-markers.}
		\label{fig_mag_response_vsize}
	\end{center}
\end{figure}
Now to see if the magnetic responses has any dependence on the system size, size-dependent simulations were performed for different system sizes. The results, for both the ABA and AAB configurations, are shown in Figure \ref{fig_mag_response_vsize}. Here, for both configurations, two sets were chosen: $J_{AA}/J_{BB}=0.1$ and $J_{AB}/J_{BB}=-0.3$ where compensation is present and $J_{AA}/J_{BB}=0.9$ and $J_{AB}/J_{BB}=-0.3$ where compensation is absent. The observation is, the region where the compensation point lies, has no detectable size dependence in this resolution. But the critical points shift with changes in lattice size, indicating a possible finite-size scaling (FSS) behaviour. To find out precise estimates for $T_{comp}$ and $T_{crit}$ as functions of the
Hamiltonian parameters, the methods employed are discussed in detail, in Sections C and D, respectively.
\vspace{10pt}
\begin{center}{\Large \textbf {B. Morphological studies :}}\end{center}
\vspace{10pt}

\begin{figure}[!htb]
	\begin{center}
		\begin{tabular}{c}
			(a)
			\resizebox{5.5cm}{!}{\includegraphics[angle=0]{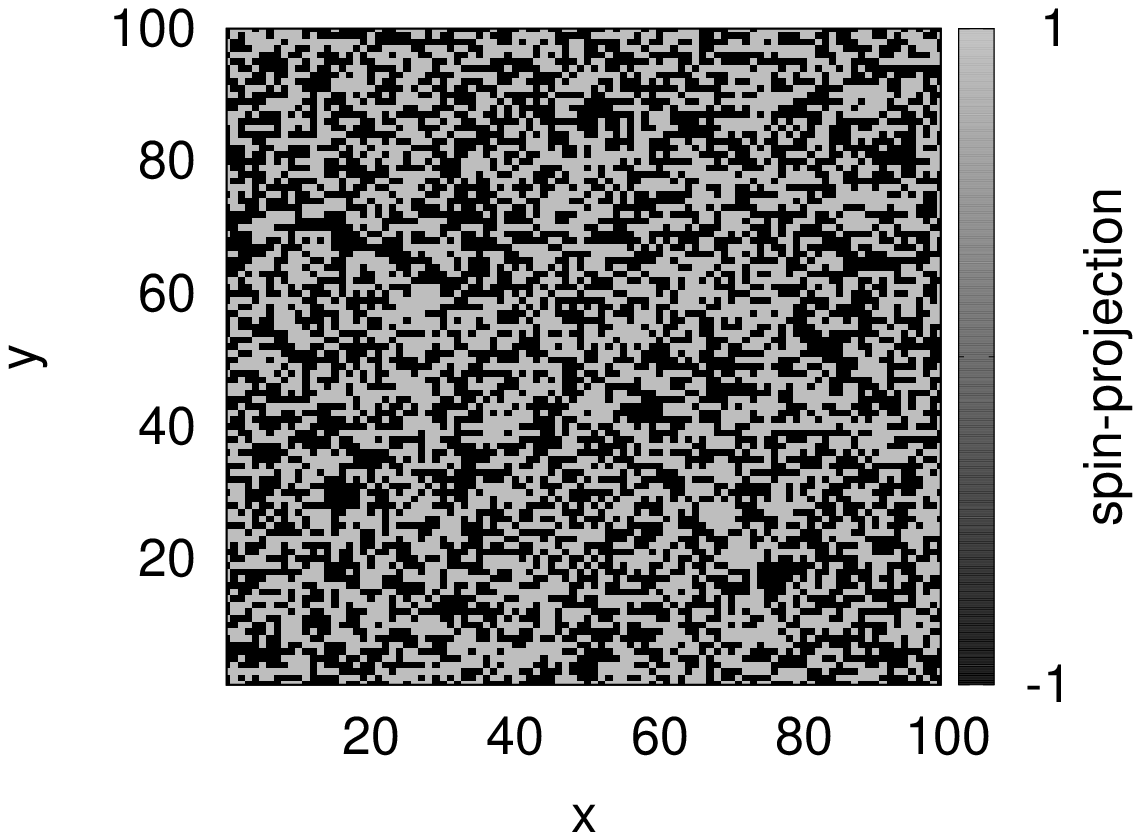}}
			(b)
			\resizebox{5.5cm}{!}{\includegraphics[angle=0]{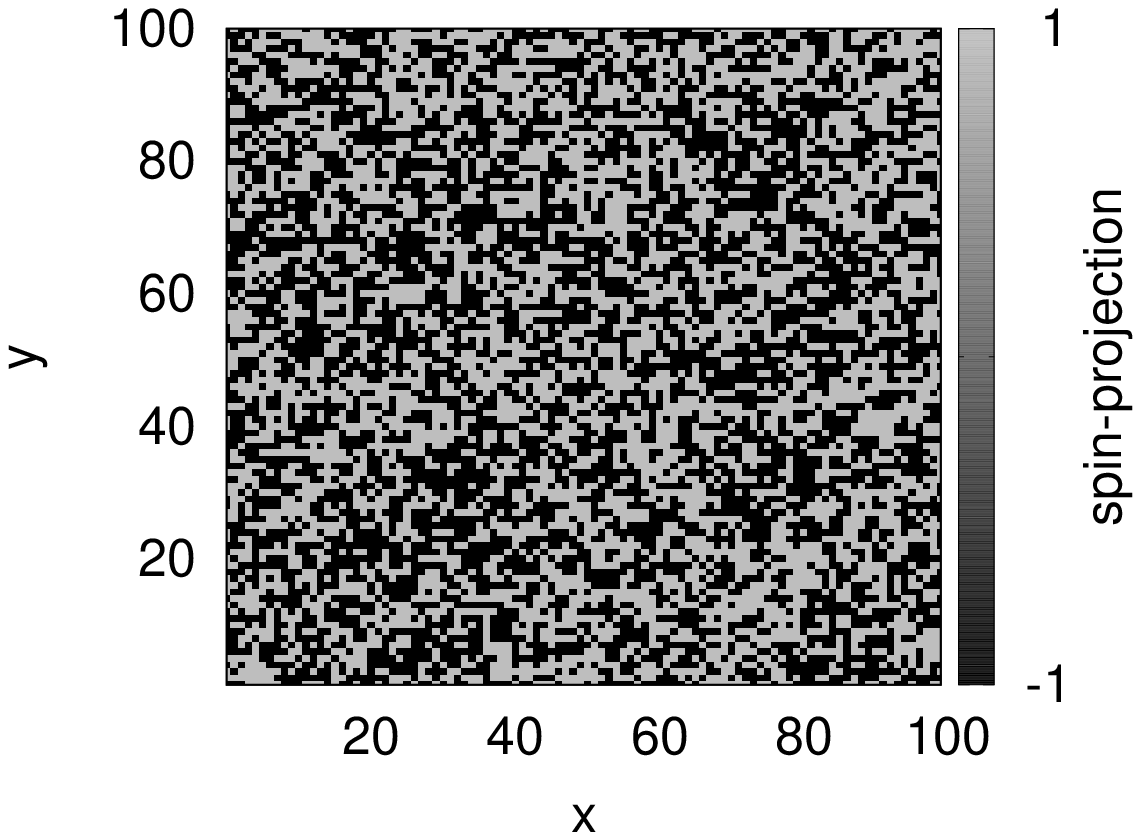}}
			(c)
			\resizebox{5.5cm}{!}{\includegraphics[angle=0]{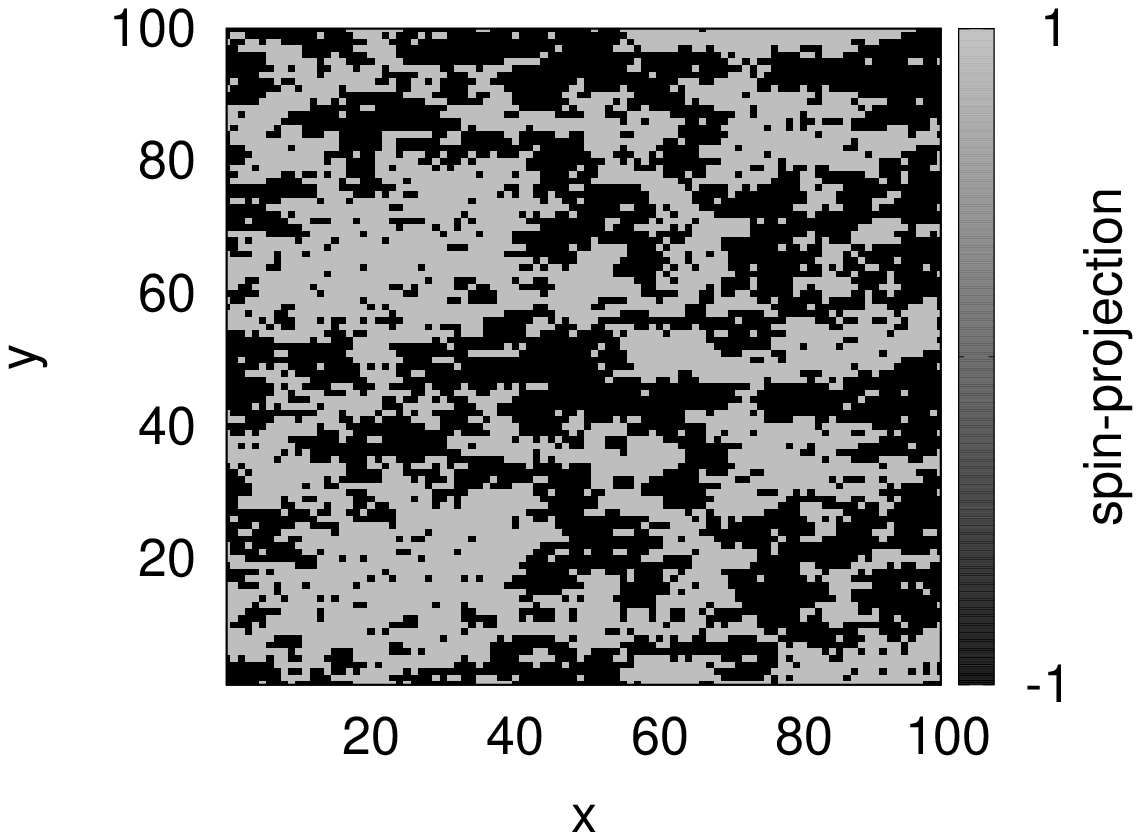}}
		\end{tabular}
		\caption{Morphology of (a) Top layer, (b) Mid layer and (c) Bottom layer for the AAB stacking ($J_{AA}/J_{BB}=0.4$ and $J_{AB}/J_{BB}=-0.3$) at (just below $T_{crit}$) $T=3.85$ with $M_{t}=-3.40\times10^{-3}$, $M_{m}=-1.33\times10^{-2}$, $M_{b}=5.74\times10^{-2}$ and $M=-1.36\times10^{-2}$}		
		\label{fig_aab_morpho1}
	\end{center}
\end{figure}

\begin{figure}[!htb]
	\begin{center}
		\begin{tabular}{c}
			(a)
			\resizebox{5.5cm}{!}{\includegraphics[angle=0]{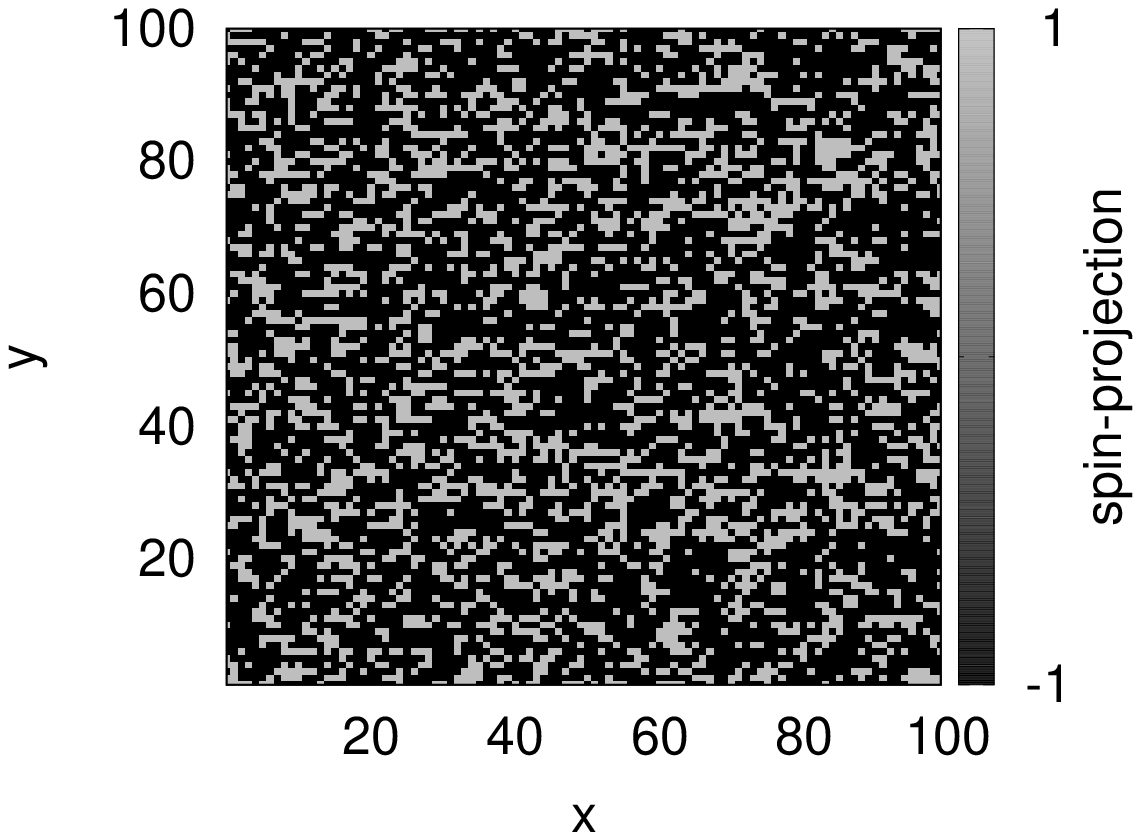}}
			(b)
			\resizebox{5.5cm}{!}{\includegraphics[angle=0]{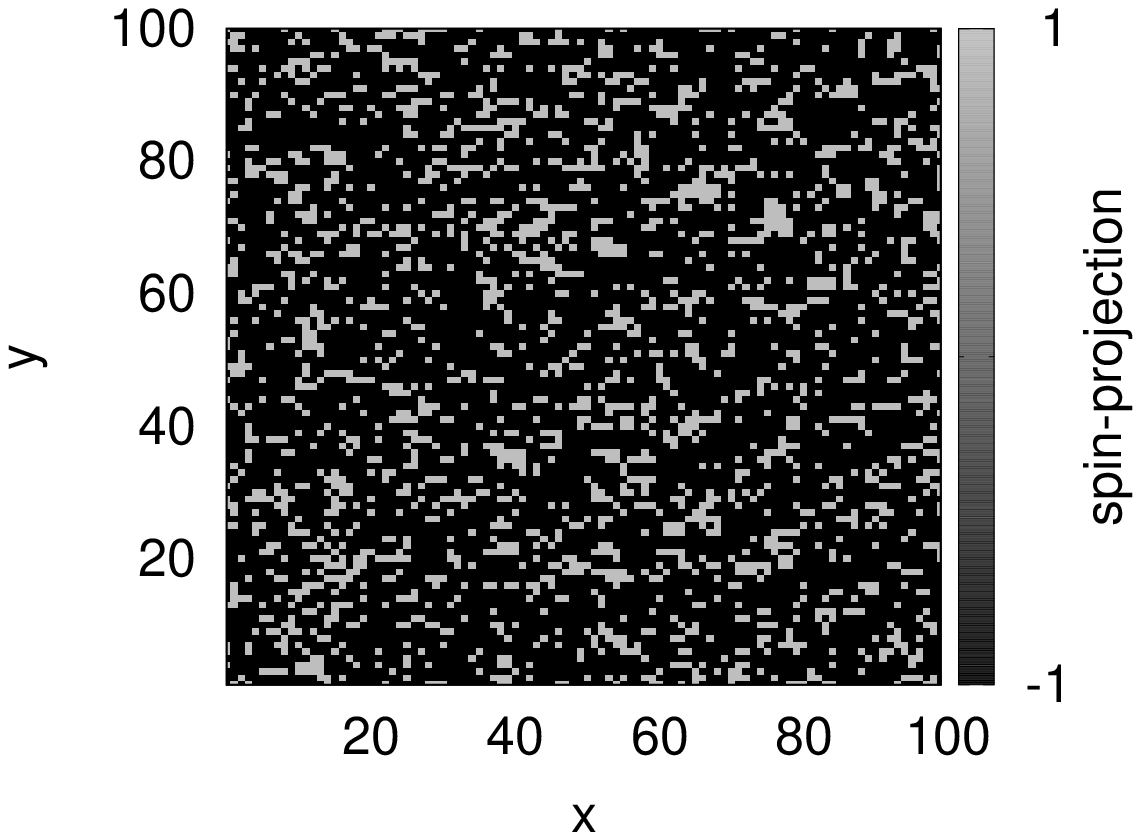}}
			(c)
			\resizebox{5.5cm}{!}{\includegraphics[angle=0]{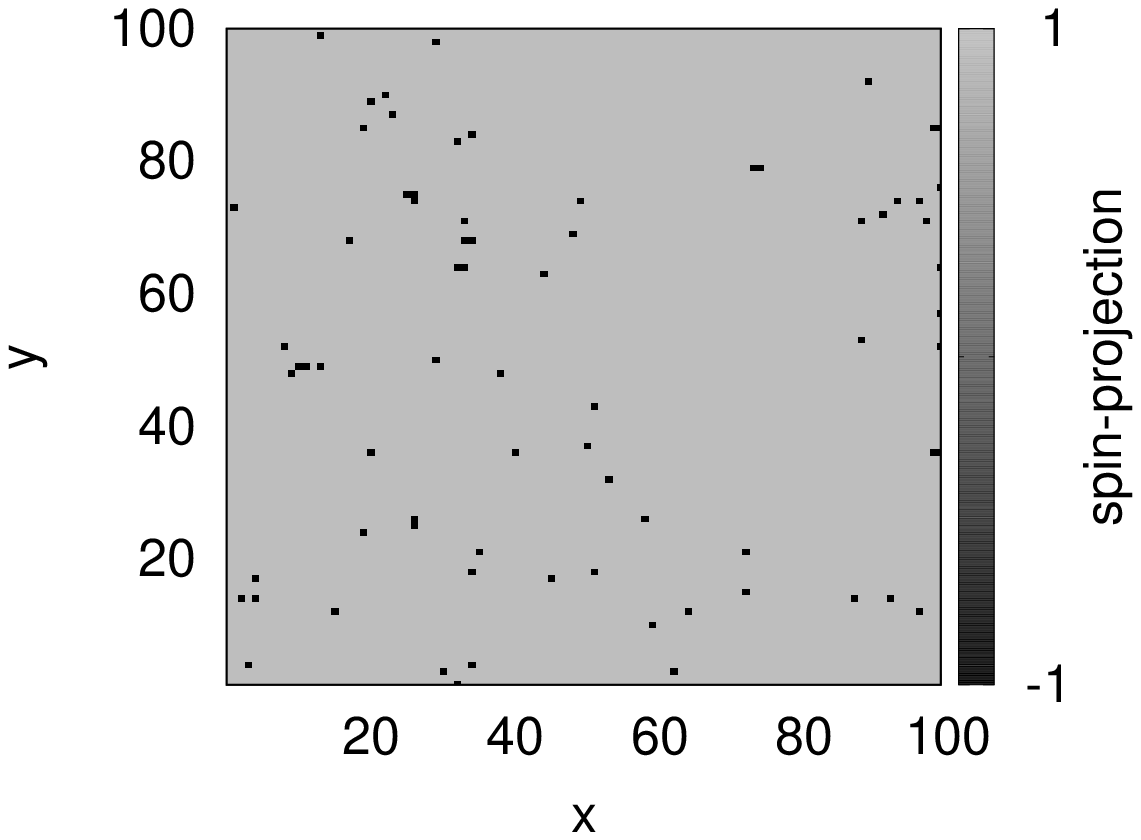}}
		\end{tabular}
		\caption{Morphology of (a) Top layer, (b) Mid layer and (c) Bottom layer for the AAB stacking ($J_{AA}/J_{BB}=0.4$ and $J_{AB}/J_{BB}=-0.3$) at $T=2.40$ with $M_{t}=-0.368$, $M_{m}=-0.567$, $M_{b}=0.984$ and $M=1.66\times 10^{-2}$.}		
		\label{fig_aab_morpho2}
	\end{center}
\end{figure}

\begin{figure}[!htb]
	\begin{center}
		\begin{tabular}{c}
			(a)
			\resizebox{5.5cm}{!}{\includegraphics[angle=0]{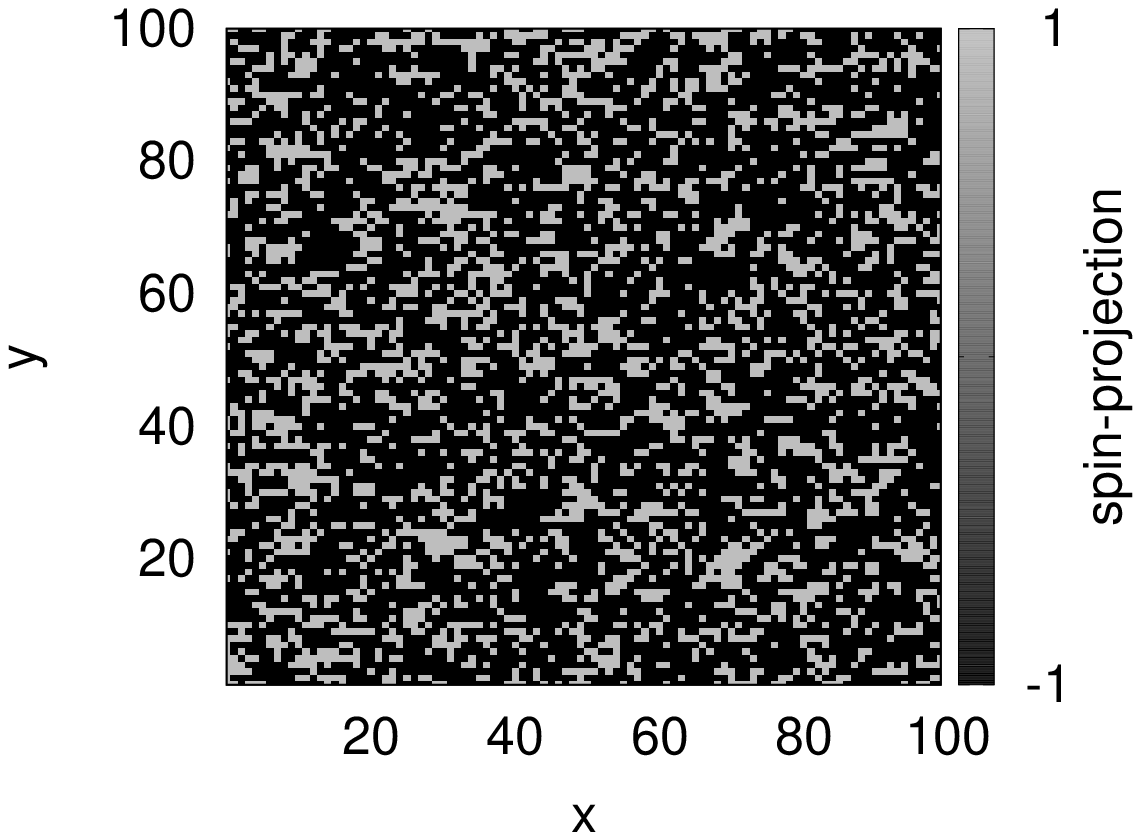}}
			(b)
			\resizebox{5.5cm}{!}{\includegraphics[angle=0]{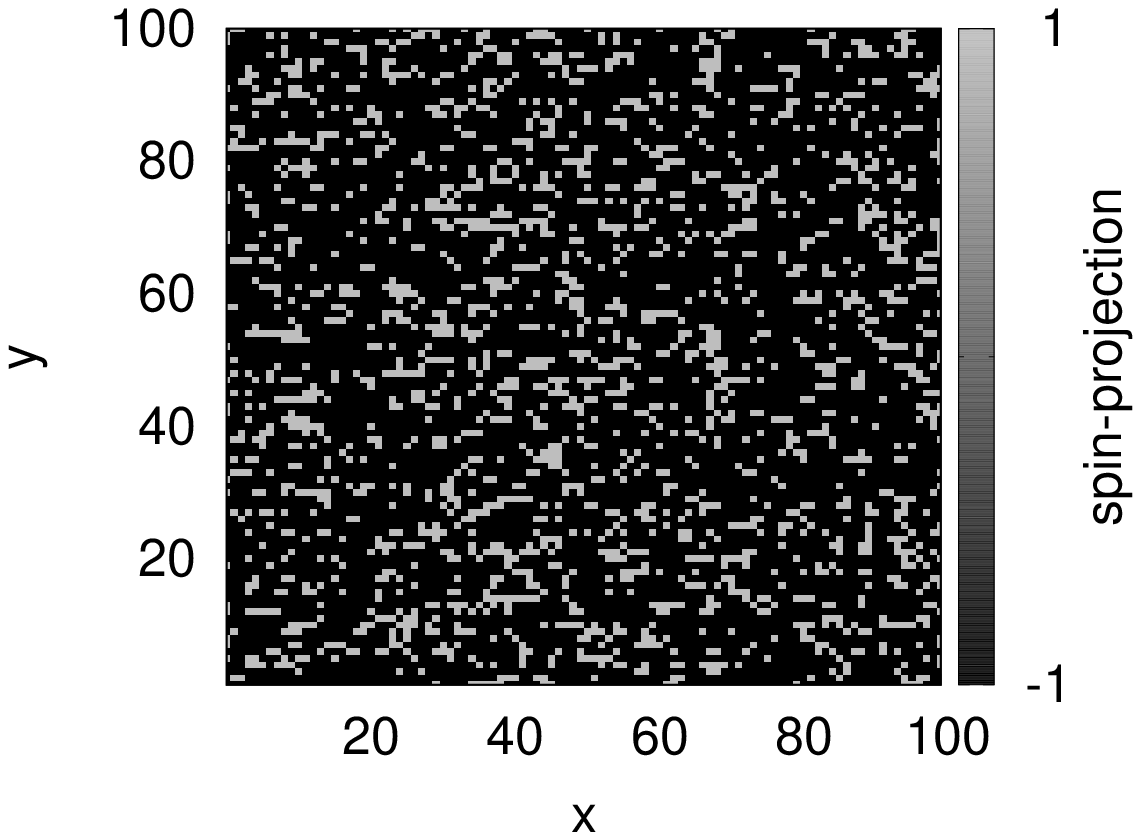}}
			(c)
			\resizebox{5.5cm}{!}{\includegraphics[angle=0]{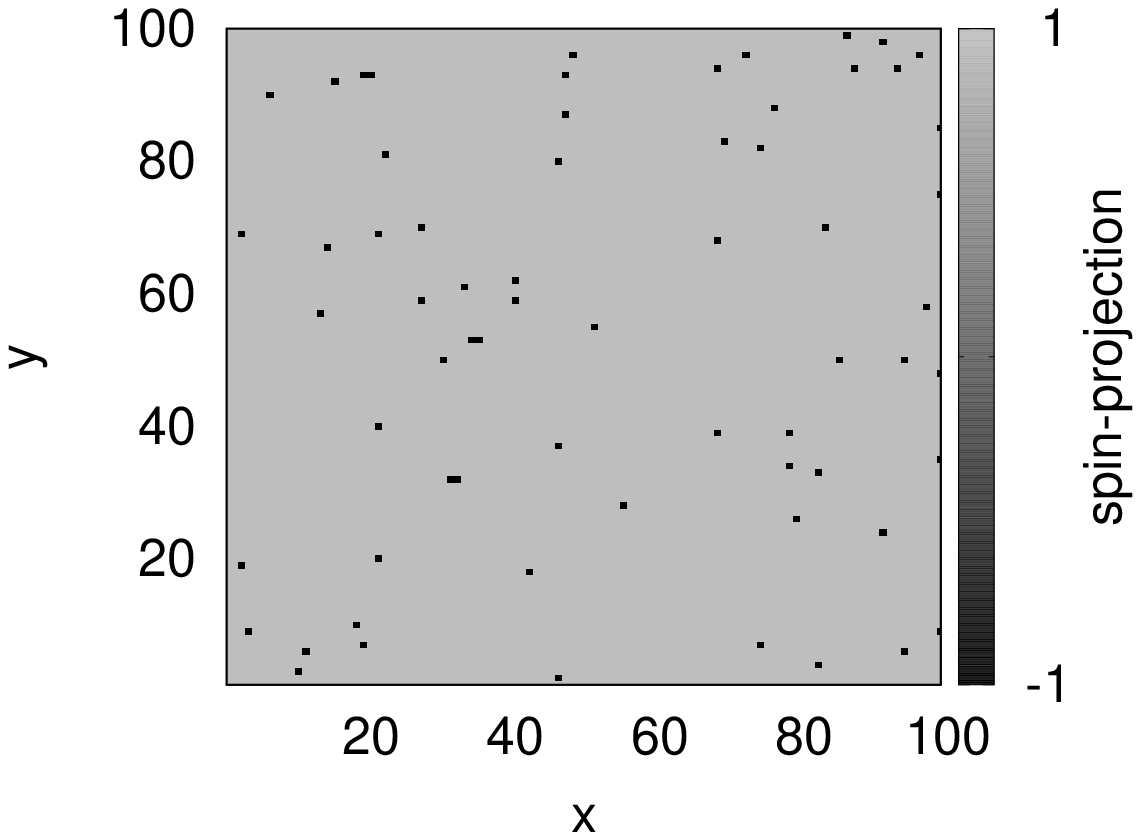}}
		\end{tabular}
		\caption{Morphology of (a) Top layer, (b) Mid layer and (c) Bottom layer for the AAB stacking ($J_{AA}/J_{BB}=0.4$ and $J_{AB}/J_{BB}=-0.3$) at $T=2.35$ with $M_{t}=-0.408$, $M_{m}=-0.603$, $M_{b}=0.986$ and $M=-8.04\times 10^{-3}$.}		
		\label{fig_aab_morpho3}
	\end{center}
\end{figure}

The lattice morphologies around $T_{comp}$ and $T_{crit}$ are investigated which is not very common in similar types of simulational studies. Such studies of the systems throw light upon how magnetic ordering develops below the critical temperatre. For both the AAB and ABA configurations, same interaction strengths $(J_{AA} /J_{BB} = 0.4; J_{AA} /J_{BB} = -0.3)$ were chosen for morphological stdies, such that the compensation effect is present in both of these configurations. Tiny white squares denote spin projections, $S^{z}=+1$ and tiny black squares represent spin projections, $S^{z}=-1$.\\
\indent For AAB stacking, Figure \ref{fig_aab_morpho1} is the spin density maps of the layers at just below the critical temperature, $T_{crit}$ and Figures \ref{fig_aab_morpho2} and \ref{fig_aab_morpho3} are at immediate higher and lower temperatures than the compensation temperature, $T_{comp}$, respectively. The sublattice magnetizations and the average magnetisation, in the vicinity of $T_{crit}$ are practically vanishing. However, it is interesting to note the value of the magnetization of B-layer at this temperature ($\sim$ critical temperature). Magnetizations are in the order of $10^{-3}$ in the top A layer and $10^{-2}$ in the mid A layer while on the other hand the B layer has the magnetization in the order of $10^{-2}$ but $5$ times of higher than that of the mid layer. It is evident that at just below $T_{crit}$, top and mid $A$-layers (with moderate in-plane coupling strengths) is occupied by almost an equal amount of up and down spins. But the bottom $B$-layer, with dominant in-plane coupling strength has already started to develop detectable magnetic order and its magnitude for magnetization is far greater than that of the other two $A$-layers. This can be understood from the larger clusters forming in the morphology of B-layer, at $T_{crit}$ as shown in Figure \ref{fig_aab_morpho1}(c), leading to higher value of magnetisation than the rest. The atoms in the mid $A$-layer is thus influenced by antiferromagnetic coupling from the atoms of the bottom $B$-layer. That is why the magnitude of mid $A$-layer magnetization is greater than that of the top $A$-layer.
\\\indent Now in the vicinity of $T_{comp}$ (Figures \ref{fig_aab_morpho2} and \ref{fig_aab_morpho3}), the magnetic clusters grow larger in size (for $T_{comp}<T_{crit}$). In this case, both the A layers are dominated by down spins whereas the B layer is nearly saturated by up spins. So non-zero values of layered magnetizations are seen around $T_{comp}$. The difference in
the size of the spin clusters creates unequal values of layered magnetizations. But the total magnetization of the bulk becomes zero leading to the phenomenon of compensation. From the configurational details, the conditions of compensation for the AAB type system can be written as, 
\begin{eqnarray}
|M_{b}| &=& |M_{t}+M_{m}|\\
sgn(M_{t}) = -sgn(M_{b}) &;& sgn(M_{m}) = -sgn(M_{b})
\end{eqnarray}

\begin{figure}[!htb]
	\begin{center}
		\begin{tabular}{c}
			(a)
			\resizebox{5.5cm}{!}{\includegraphics[angle=0]{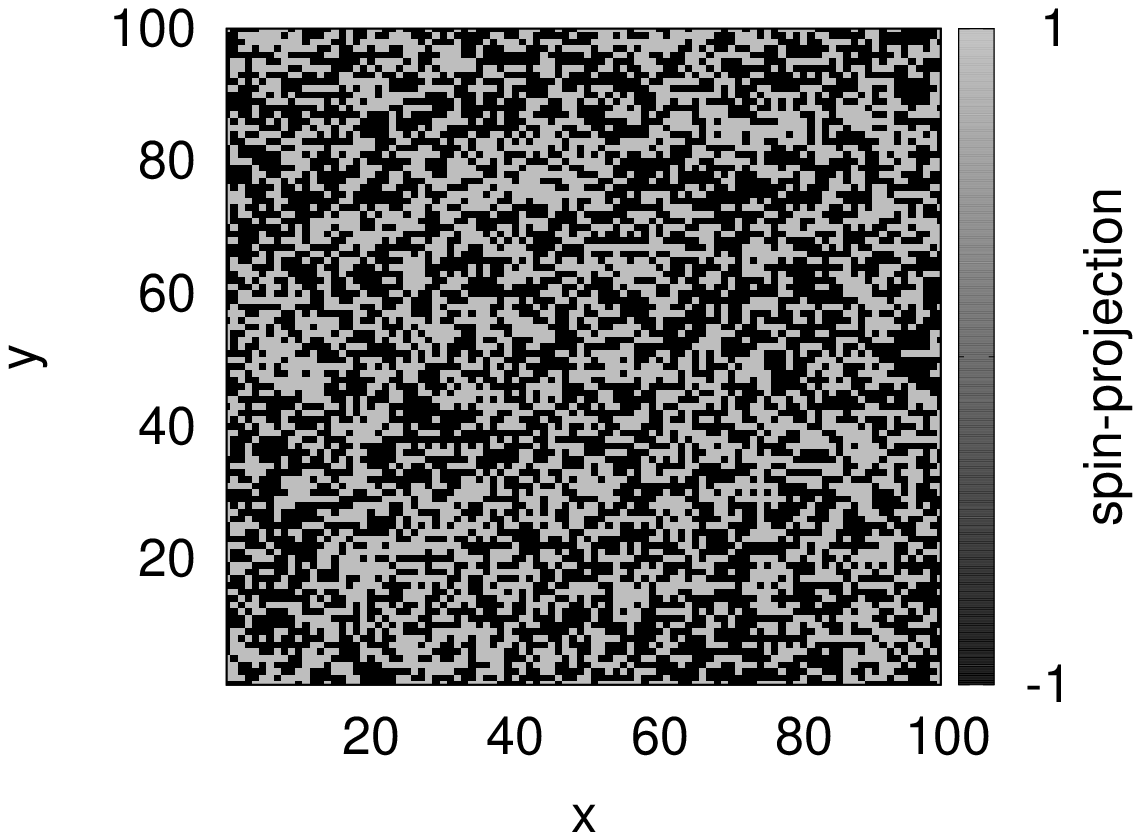}}
			(b)
			\resizebox{5.5cm}{!}{\includegraphics[angle=0]{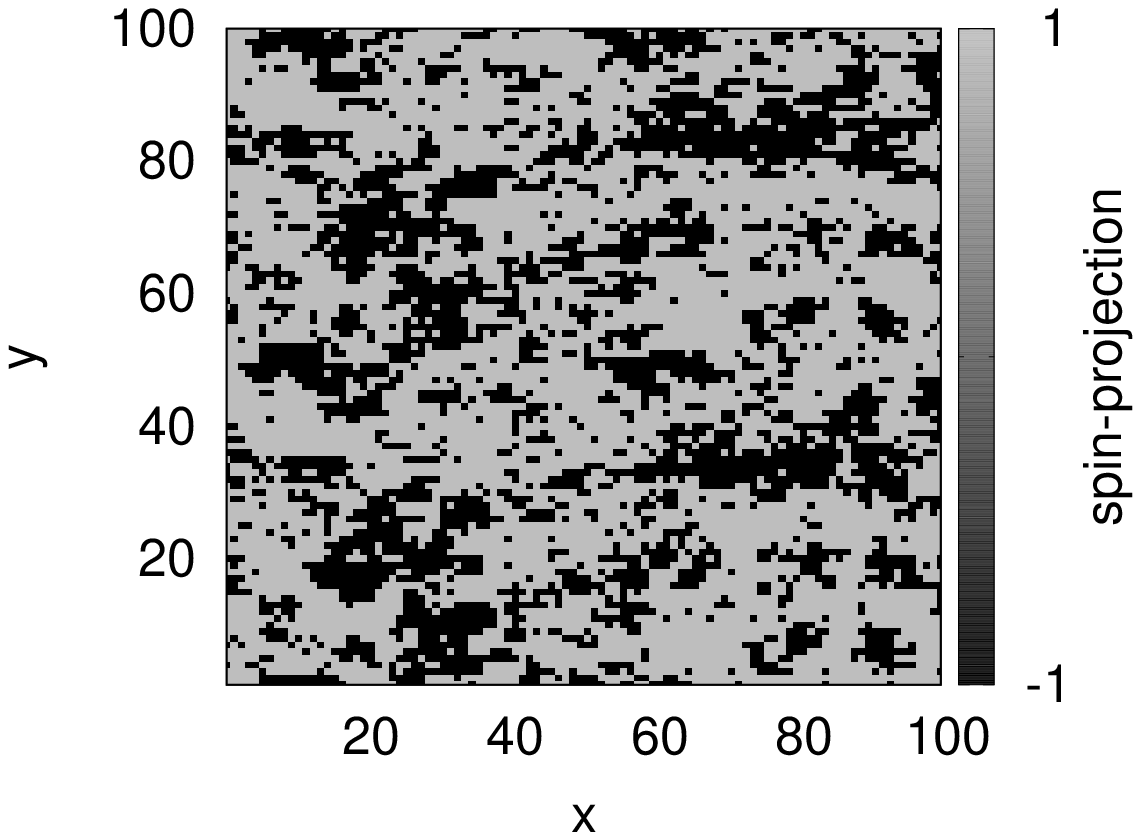}}
			(c)
			\resizebox{5.5cm}{!}{\includegraphics[angle=0]{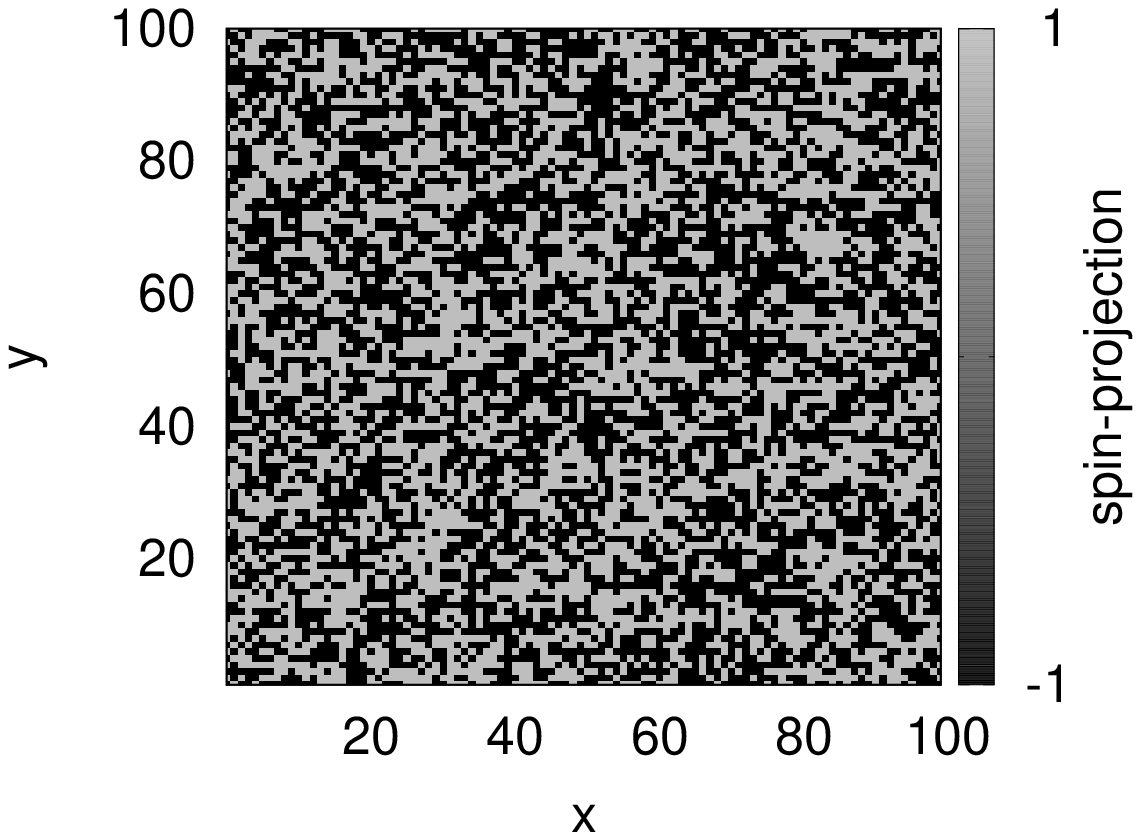}}
		\end{tabular}
		\caption{Morphology of (a) Top layer, (b) Mid layer and (c) Bottom layer for the ABA stacking ($J_{AA}/J_{BB}=0.4$ and $J_{AB}/J_{BB}=-0.3$) at $T_{crit}=3.90$ with $M_{t}=-7.75\times10^{-3}$, $M_{m}=4.28\times10^{-2}$, $M_{b}=-5.42\times10^{-3}$.} and $M=9.87\times 10^{-3}$	
		\label{fig_aba_morpho1}
	\end{center}
\end{figure}

\begin{figure}[!htb]
	\begin{center}
		\begin{tabular}{c}
			(a)
			\resizebox{5.5cm}{!}{\includegraphics[angle=0]{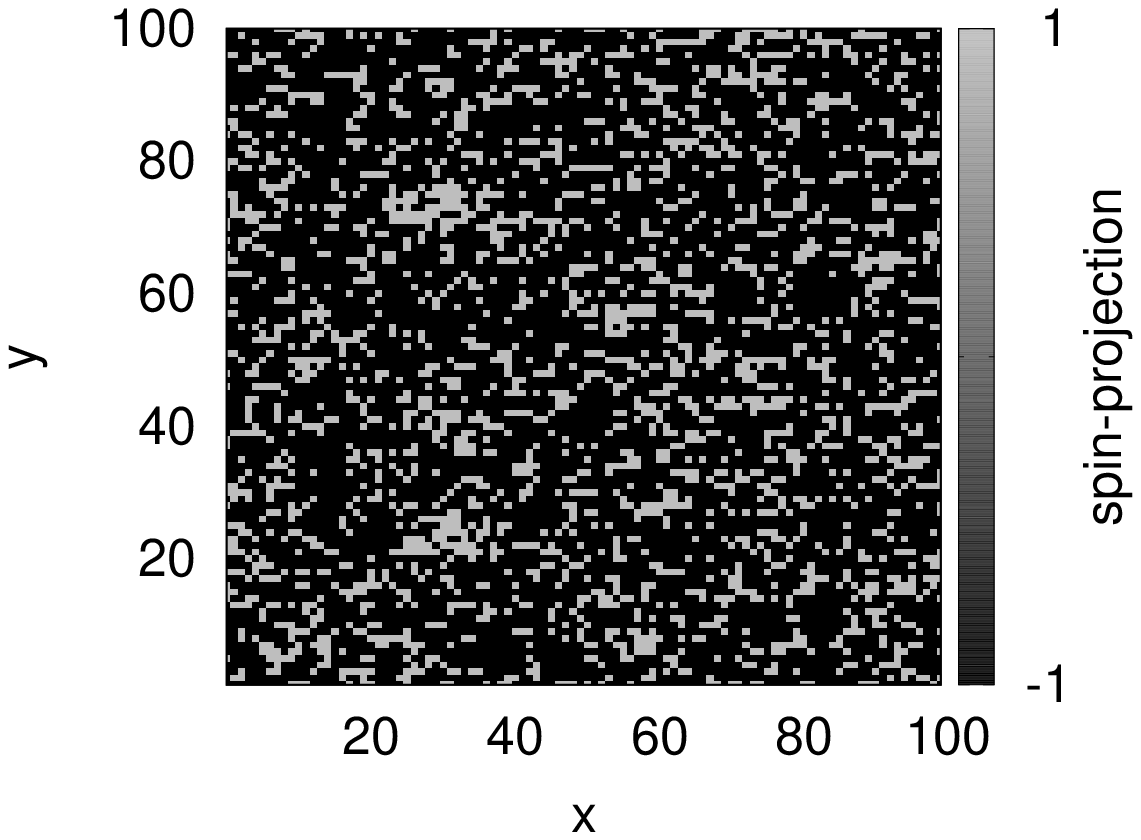}}
			(b)
			\resizebox{5.5cm}{!}{\includegraphics[angle=0]{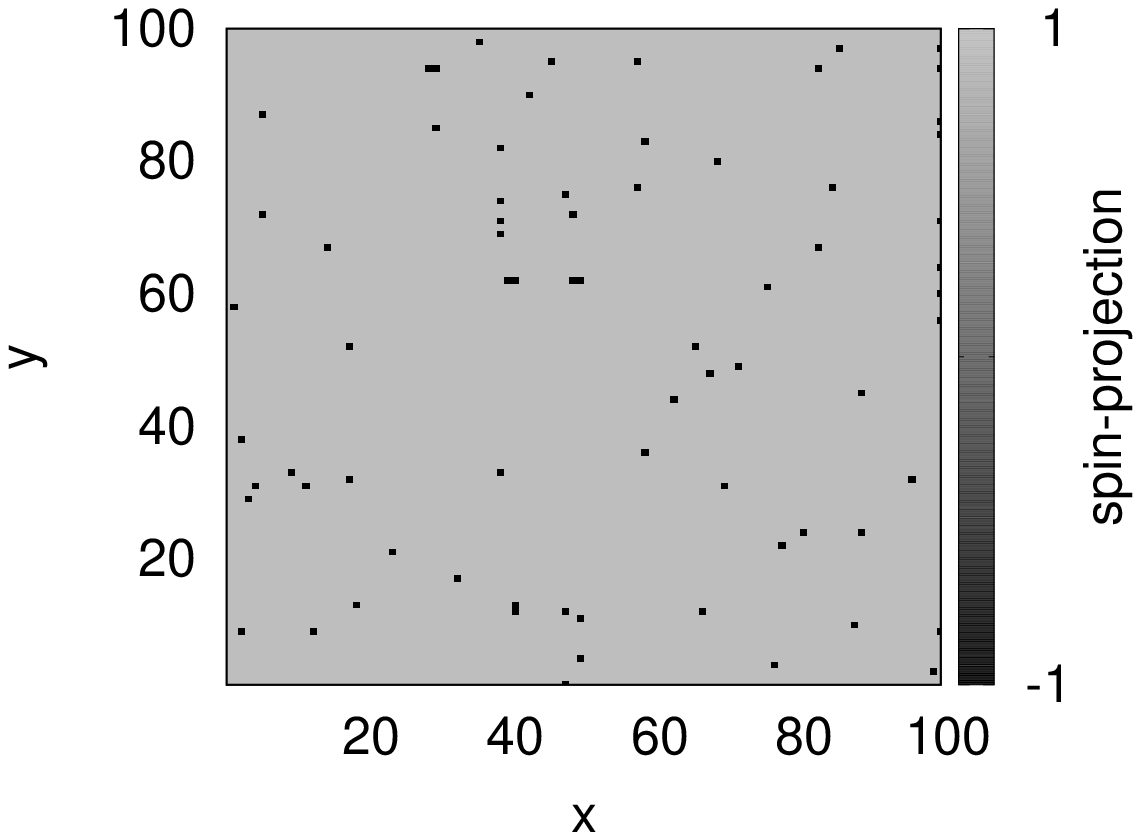}}
			(c)
			\resizebox{5.5cm}{!}{\includegraphics[angle=0]{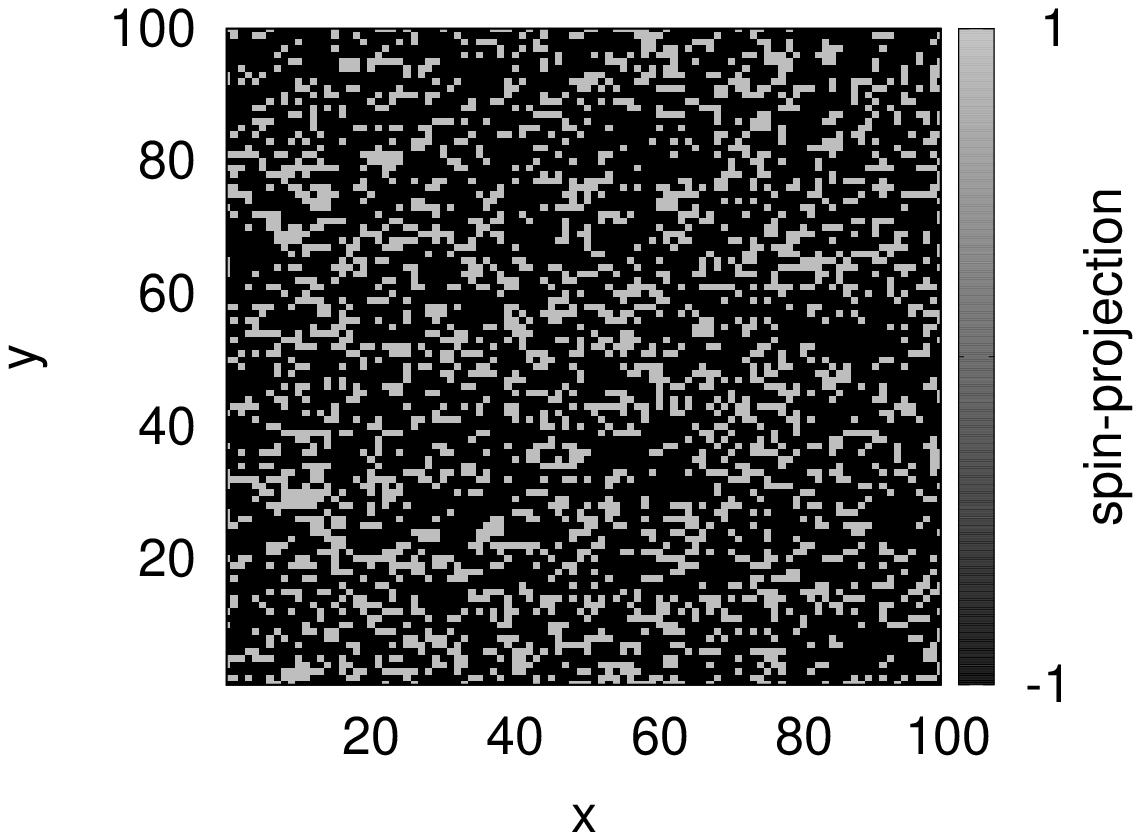}}
		\end{tabular}
		\caption{Morphology of (a) Top layer, (b) Mid layer and (c) Bottom layer for the ABA stacking ($J_{AA}/J_{BB}=0.4$ and $J_{AB}/J_{BB}=-0.3$) at $T=2.40$ with $M_{t}=-0.478$, $M_{m}=0.985$, $M_{b}=-0.479$ and $M=9.65\times 10^{-3}$.}		
		\label{fig_aba_morpho2}
	\end{center}
\end{figure}

\begin{figure}[!htb]
	\begin{center}
		\begin{tabular}{c}
			(a)
			\resizebox{5.5cm}{!}{\includegraphics[angle=0]{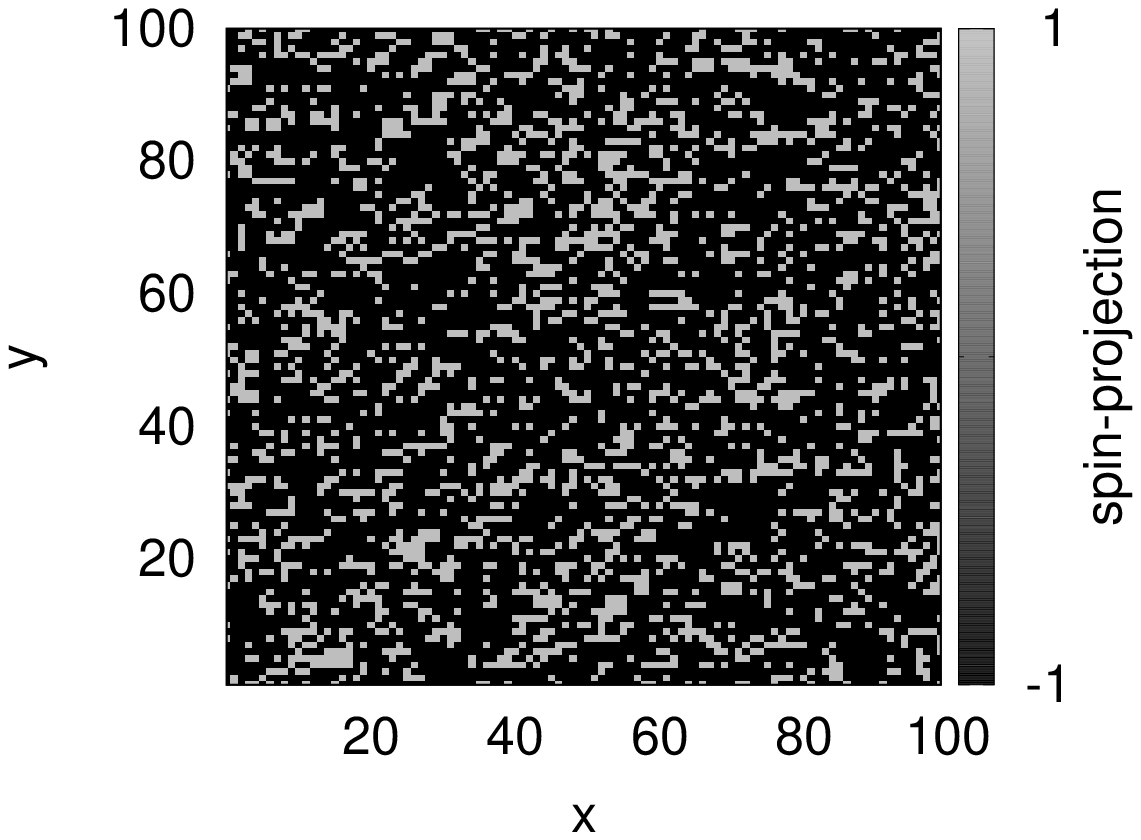}}
			(b)
			\resizebox{5.5cm}{!}{\includegraphics[angle=0]{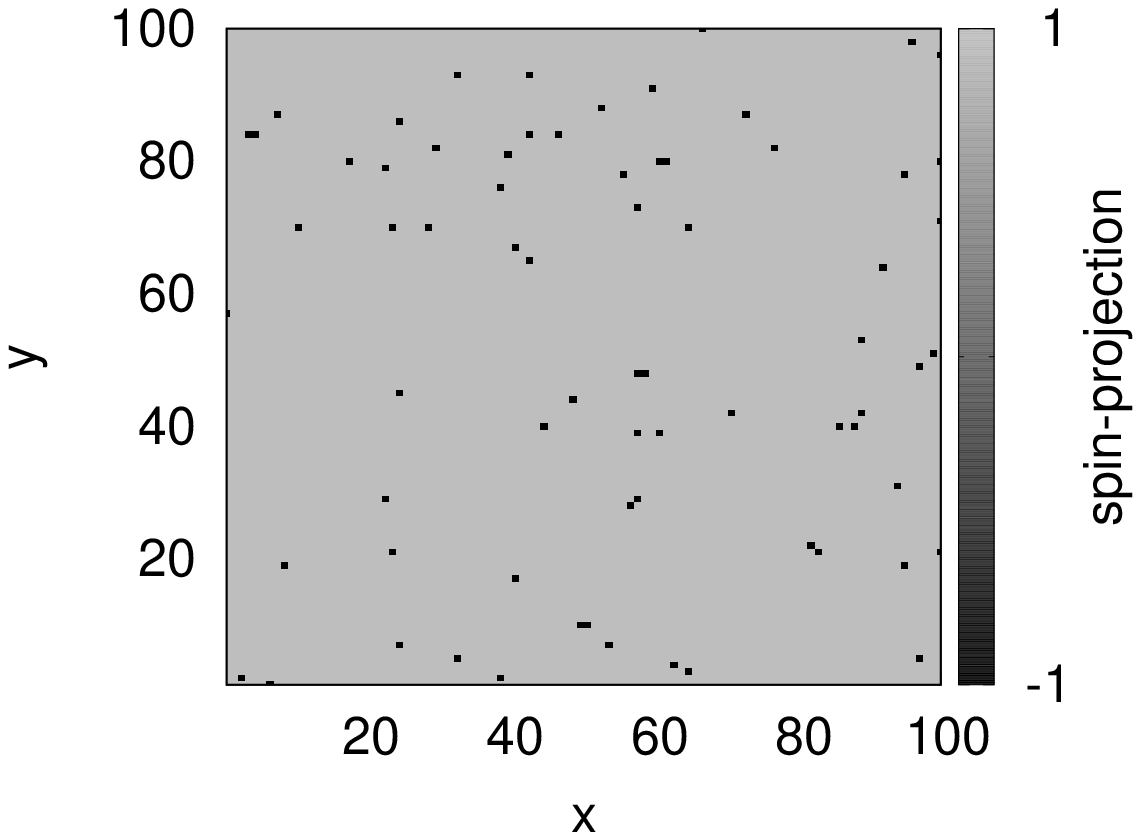}}
			(c)
			\resizebox{5.5cm}{!}{\includegraphics[angle=0]{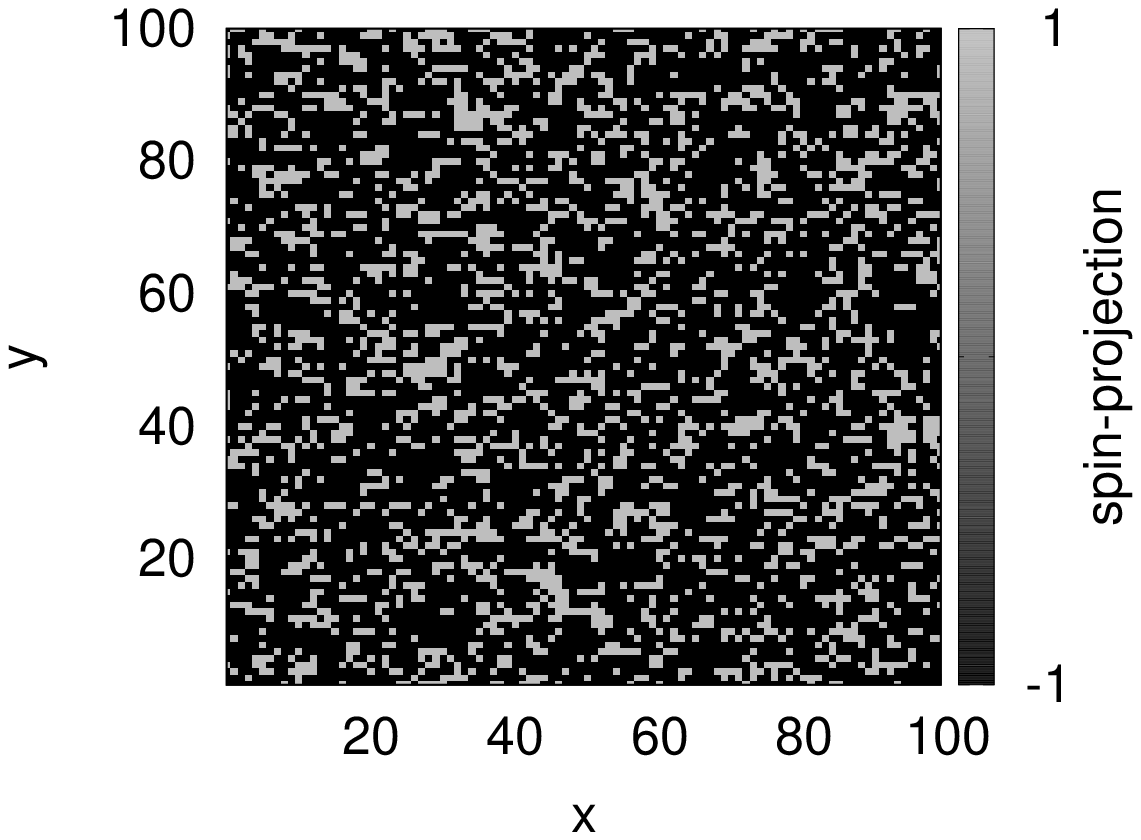}}
		\end{tabular}
		\caption{Morphology of (a) Top layer, (b) Mid layer and (c) Bottom layer for the ABA stacking ($J_{AA}/J_{BB}=0.4$ and $J_{AB}/J_{BB}=-0.3$) at $T=2.35$ with $M_{t}=-0.504$, $M_{m}=0.988$, $M_{b}=-0.505$ and $M=-6.99\times 10^{-3}$.}		
		\label{fig_aba_morpho3}
	\end{center}
\end{figure}

For ABA stacking, Figure \ref{fig_aba_morpho1} contains the density maps of the layers at critical temperature, $T_{crit}$ and Figures \ref{fig_aba_morpho2} and \ref{fig_aba_morpho3} contain spin density maps at immediate higher and lower temperatures of compensation temperature, $T_{comp}$, respectively. Like AAB type, at $T_{crit}$ for ABA system, every layer is occupied by almost equal up and down spins leading to vanishing sublattice and consequently vanishing average magnetisation. Here the magnetizations are in the order of $10^{-3}$ in the top and bottom A layers and $10^{-2}$ in the mid, B layer. The larger clusters, in the morphology of B-layer, at $T_{crit}$ as shown in Figure \ref{fig_aba_morpho1}(b), leads to such higher value of magnetisation in the mid layer.

Now in the vicinity of $T_{comp}$ (Figures \ref{fig_aba_morpho2} and \ref{fig_aba_morpho3}), spin clusters get bigger in size and layers are going towards saturation (for $T_{comp}<T_{crit}$). Again, both the A layers are dominated by down spins while the B layer is nearly saturated by up spins, leading to non-zero values of layered magnetizations at $T_{comp}$. The difference in the size of the spin clusters is clearly visible in the figures. From the configurational details of the ABA type system, the conditions of compensation is written as, 
\begin{eqnarray}
|M_{m}| &=& |M_{t}+M_{b}|\\
sgn(M_{t}) = -sgn(M_{m}) &;& sgn(M_{b}) = -sgn(M_{m})
\end{eqnarray}

\begin{figure}[!htb]
	\begin{center}
		\begin{tabular}{c}
			\resizebox{10cm}{!}{\includegraphics[angle=0]{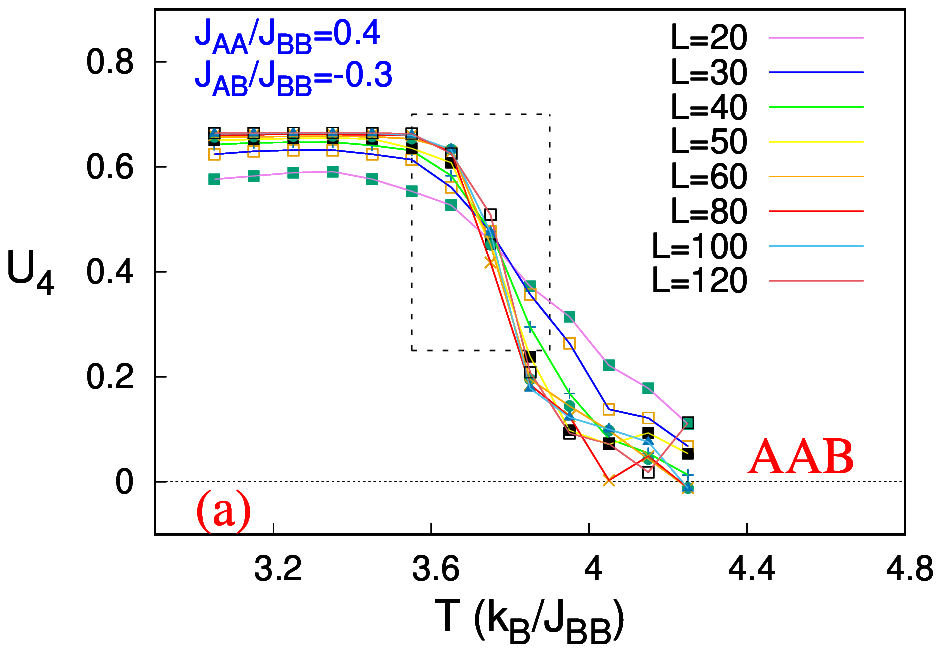}}
			\resizebox{10cm}{!}{\includegraphics[angle=0]{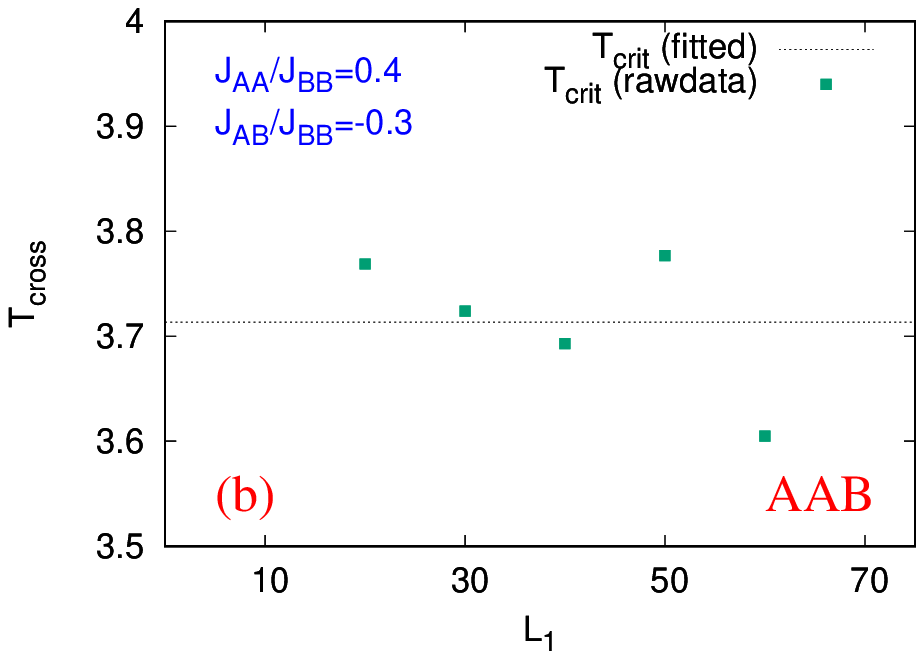}}\\
			\resizebox{10cm}{!}{\includegraphics[angle=0]{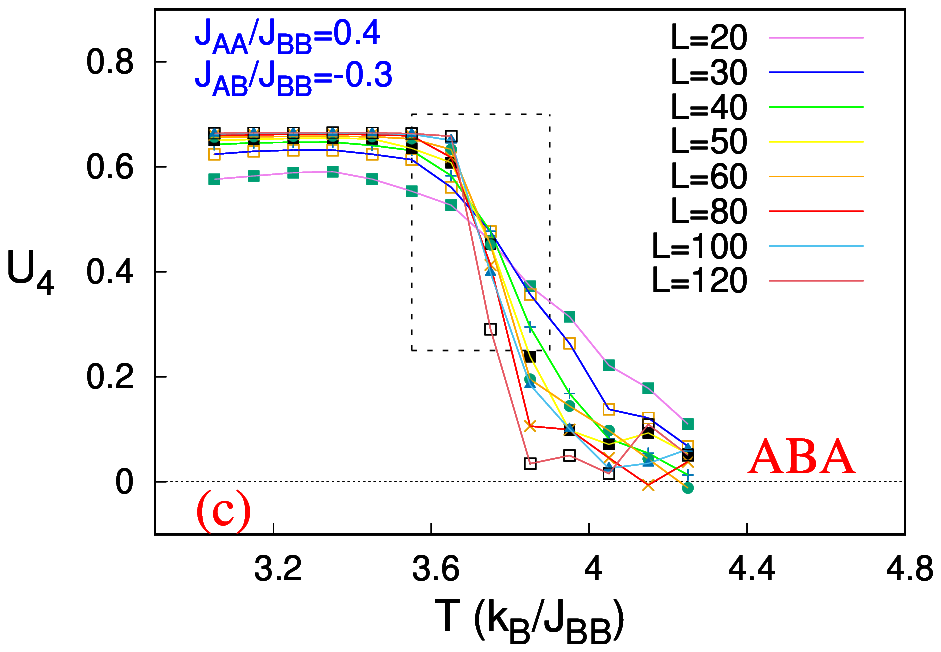}}
			\resizebox{10cm}{!}{\includegraphics[angle=0]{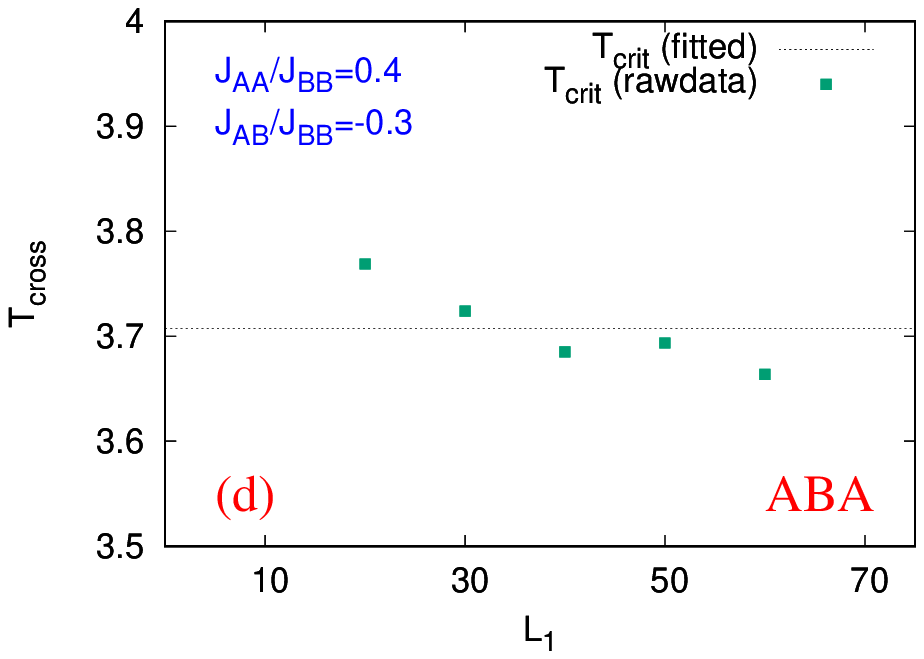}}
		\end{tabular}
		\caption{ (Colour Online) Plot of fourth order magnetization cumulant, $U_{4}$ as a function of dimensionless Temperature, for (a) AAB (c) ABA configurations and fitting of data for (b) AAB (d) ABA configuartions ($J_{AA}/J_{BB}=0.4$ and $J_{AB}/J_{BB}=-0.3$ for both). The dimensionless critical temperatures, came out to be $3.7134\pm0.0312$ for AAB and $3.7070\pm0.0182$ for ABA configuration, best values of which are marked by dashed lines in (b) and (d)}		
		\label{fig_crit_simfit}
	\end{center}
\end{figure}

The spin density plots establish the fact that the $T_{comp}$ and $T_{crit}$ are two fundamentally different points because of their different lattice morphologies. The formation of asymmetric spin clusters at $T_{comp}$ is responsible for compensation phenomenon as the layer with highest in-plane exchange coupling hosts largest ordered spin cluster and its magnetization gets cancelled by the other two. The values of layered magnetizations in each case show, usual mathematical relations between sublattice magnetizations, at $T_{comp}$, are also obeyed. 
\vspace{10pt}
\begin{center}{\Large \textbf {C. Evaluation of Compensation and Critical temperatures :}}\end{center}
\vspace{10pt}
\indent \textit{Compensation temperature} is that temperature, where the system, as a whole, shows $<M>=0$ while individual layers still remain magnetized (i.e. $M_{q}\neq 0$), as seen in Figs. \ref{fig_aab_morpho2}, \ref{fig_aab_morpho3}, \ref{fig_aba_morpho2} and \ref{fig_aba_morpho3}. For both, AAB and ABA configurations, to find this temperature for all the different combinations of coupling strengths, simulations were performed for a few equally spaced temperatures (interval of $0.004$) around \textit{quasi} $T_{comp}$ (obtained from the simulations of Figure \ref{fig_mag_response}) and the average magnetization values were plotted against temperature, for different system sizes (ranging from $L=40$ to $L=120$). Then, using linear interpolation, the temperature coordinate of the point, where $<M>$ crosses the zero magnetization line was found, and these were plotted as functions of system size, $L$. \\
\indent It is seen that different $T_{comp}(L)$ are confined within a narrow band. Figures \ref{fig_comp_simfit}(b) and \ref{fig_comp_simfit}(d) shows the size dependence of the compensation temperature estimates obtained from the linear interpolations of plots in Figures \ref{fig_comp_simfit}(a) and \ref{fig_comp_simfit}(c). After a certain value of $L$, the compensation temperature gets trapped within a narrow range. Next up, the values of $T_{comp}$ were fitted, according to Equation (\ref{eq_tcomp}):
\begin{equation}
T_{comp}(L)=a
\label{eq_tcomp}
\end{equation}
where $a$ is a constant. 
\\\indent For $L\geq70$, the values approximately converge to the fitted value. But the system sizes, which should be ignored while finally fitting the data, is determined from the well known \textit{reduced chi-squared} ($\chi^{2}/n_{DOF}$) \footnote{Number of degrees of freedom in a fitting process, $n_{DOF}$: number of obtained data \textit{minus} the number of fitting parameters} values \cite{Taylor}, while fitting. Only those system sizes were considered where there are consistency in the order of $\chi^{2}/n_{DOF}$. The sizes, $L=40,50,60$ are thus excluded. The final error associated with $T_{comp}$ comes from two sources: (a) from the linear interpolation and (b) from fitting the data by Equation (\ref{eq_tcomp}). The estimate of error in (b) is obtained by Jackknife method \cite{Newman}. So both: the errors obtained in fitting process and the largest error, in finding intersections for different $L$'s, were combined for the final error estimate. Equation (\ref{eq_tcomp}) is consistent with the fact that the compensation phenomenon is not related to criticality (e.g. no power law scaling is seen). For $J_{AA}/J_{BB}=0.4$ and $J_{AA}/J_{BB}=-0.3$ , in Figures \ref{fig_comp_simfit}(a) and \ref{fig_comp_simfit}(c), for AAB and ABA systems respectively, it is shown how close range simulations behave as functions of system size, while in Figures \ref{fig_comp_simfit}(b) and \ref{fig_comp_simfit}(d), the results of the fitting procedure are shown.\\ 
\indent In Figure \ref{fig_mag_response_vsize}, possible size dependence of critical temperatures are observed for both AAB and ABA configurations. So scaling behaviour in this region may be expected. So to determine the critical temperatures precisely, the \textit{cumulant crossing technique} proposed by Binder \cite{Binder2} is employed. Binder introduced fourth-order magnetization cumulant $U_{4}$, defined by:
\begin{equation}
U_{4}=1-\dfrac{<M^{4}>}{3<M^{2}>}
\label{eq_bincum}
\end{equation}
where $M$ is the magnetization. In this approach, the magnetization cumulant of Equation [\ref{eq_bincum}], for
different lattice sizes, are plotted as a function of temperature, $T$ and all
the intersections of $U_{4}$ for any two lattice sizes, say, $L_{1}$ and $L_{2}$ of fixed ratio $b=L_{2}/L_{1}=2$ are determined, from below \cite{Binder2,Ferrenberg}. Then the estimate for critical temperature, $T_{crit}$, is found out by the arithmetic mean of all the values of intersections. The Jackknife method is used for the estimate of uncertainty in $T_{crit}$. All the intersections lie within the dashed rectangular boxes in Figures \ref{fig_crit_simfit}(a) and \ref{fig_crit_simfit}(c). The Figures \ref{fig_crit_simfit}(b) and \ref{fig_crit_simfit}(d) show the fitting process.
Spread in the values of $T_{crit}$ around the mean and multiple crossovers in the higher temperature region are due to the quality of the random number generators and finite statistics of the system \cite{Binder2}.

\begin{figure}[!htb]
	\begin{center}
		\begin{tabular}{c}
			\resizebox{10cm}{!}{\includegraphics[angle=0]{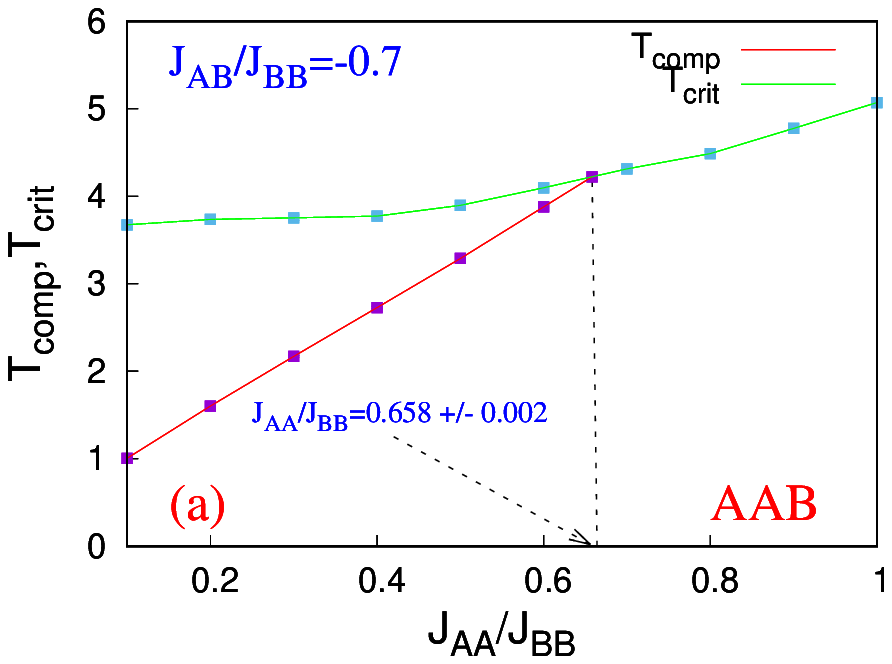}}
			\resizebox{10cm}{!}{\includegraphics[angle=0]{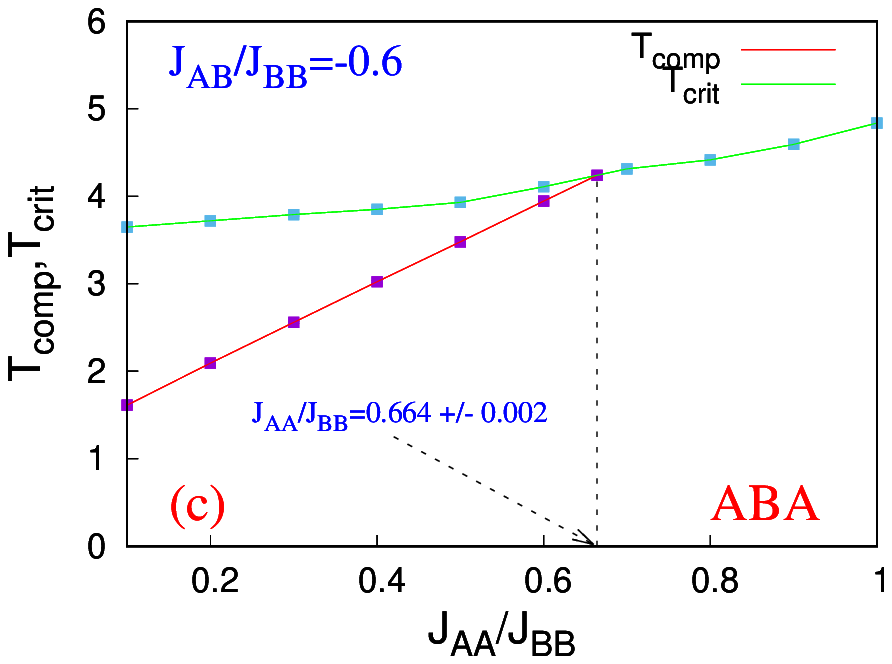}}\\
			\resizebox{10cm}{!}{\includegraphics[angle=0]{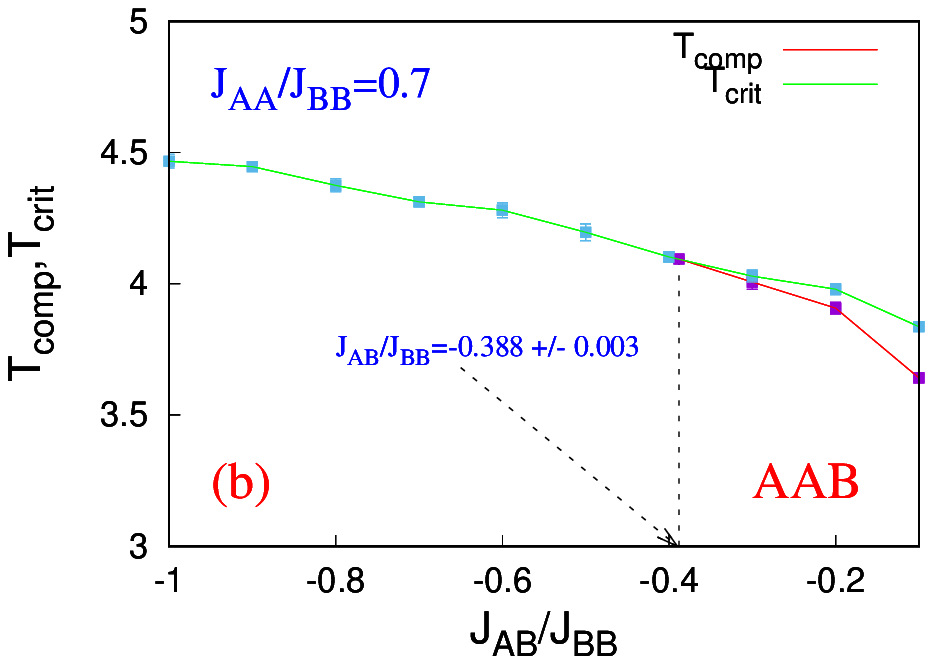}}
			\resizebox{10cm}{!}{\includegraphics[angle=0]{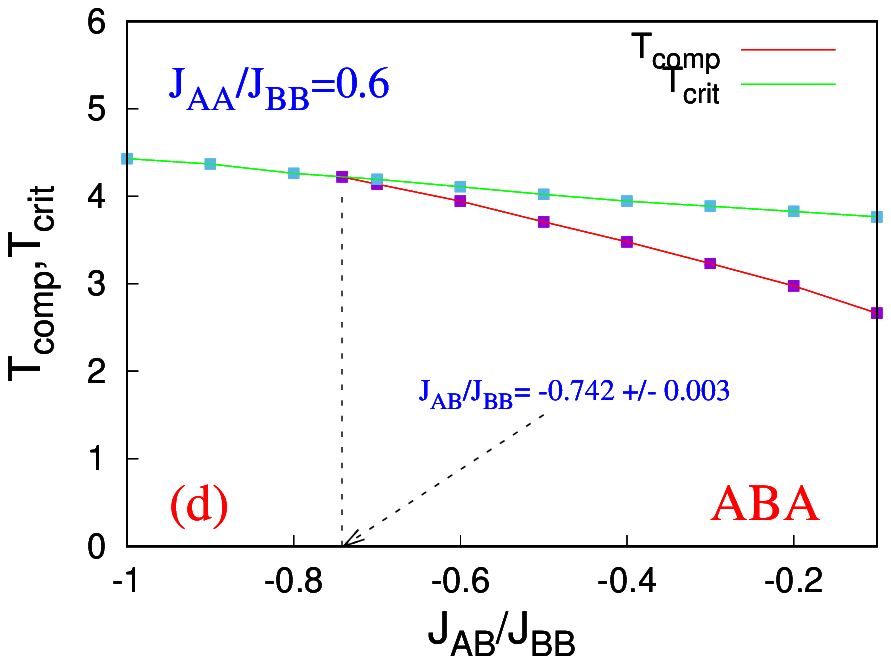}}
		\end{tabular}
			\caption{ (Colour Online) Dimensionless critical temperature $T_{crit}$ and compensation temperature $T_{comp}$: (i) as functions of $J_{AA}/J_{BB}$ for (a) AAB configuration with $J_{AB}/J_{BB}=-0.7$ (c) ABA configuration with $J_{AB}/J_{BB}=-0.6$. The dashed lines, where meet the horizontal axis, mark the value of $J_{AA}/J_{BB}$, above which no compensation is detected; and (ii) as functions of $J_{AB}/J_{BB}$ for (b) AAB configuration with $J_{AA}/J_{BB}=0.7$ (d) ABA configuration with $J_{AA}/J_{BB}=0.6$. Here the intersections of dashed lines with the horizontal axis, mark the value of $J_{AB}/J_{BB}$, below which no compensation is detected. Where the errorbars are not visible, they are smaller than the point markers.}		
	\label{fig_zeromag_bifurcation}
\end{center}
\end{figure}

It is already seen that both the critical and compensation temperatures, drift futher away from their values at low $J_{AA}/J_{BB}$, with fixed $J_{AB}/J_{BB}$ and vice-versa. But the drift of $T_{comp}$ is much rapid compared to $T_{crit}$ which results in a merger of these two temperature points, at higher up regions of relative interaction strengths. Here the dynamics of $T_{crit}$ and $T_{comp}$ is studied and one set of example is shown in Figure \ref{fig_zeromag_bifurcation}. For AAB configuration, in Figure \ref{fig_zeromag_bifurcation}(a) $J_{AB}/J_{BB}$ is kept fixed at $-0.7$ and $J_{AA}/J_{BB}$ is varied and a bifurcation of zero magnetisation curves happens at $J_{AA}/J_{BB}=0.658\pm 0.002$. In Figure \ref{fig_zeromag_bifurcation}(b), with $J_{AA}/J_{BB}=0.7$ and variable $J_{AB}/J_{BB}$, the bifurcation is observed at $J_{AB}/J_{BB}=-0.388\pm 0.003$ . For ABA configuration, in Figure \ref{fig_zeromag_bifurcation}(c) $J_{AB}/J_{BB}$ is kept fixed at $-0.6$ and $J_{AA}/J_{BB}$ is varied and a bifurcation of zero magnetisation curves is seen at $J_{AA}/J_{BB}=0.664\pm 0.002$. In Figure \ref{fig_zeromag_bifurcation}(d), with $J_{AA}/J_{BB}=0.6$ and variable $J_{AB}/J_{BB}$, the bifurcation happens at $J_{AB}/J_{BB}=-0.742\pm 0.003$ . The errors associated with finding intersections are given by the upper bounds of linear interpolation procedure \cite{Scarborough}. 
\vspace{10pt}
\begin{center}{\Large \textbf {D. Alternative description by IARRM and TICCT:}}\end{center}
\vspace{10pt}

It is always advantageous for heterostructures to have \textit{physical quantities of the bulk} which \textit{systematically varies} with parameters of the system. In the current study, investigation on \textit{Inverse absolute of reduced residual magnetisation}, IARRM and \textit{Temperature interval between Critical and Compensation temperatures}, TICCT is done to provide the alternative description for the trilayered system with trianglar sublayers. These mathematical dependences are always helpful for technological applications.

For an AAB type of system, the equations for IARRM, $\mu_{AAB}$ and TICCT, $\Delta T_{AAB}$ can be written as:
\begin{eqnarray}
\label{eq_AAB_iarrm}
\mu_{AAB} \left( \left|\dfrac{J_{AB}}{J_{BB}}\right|,\dfrac{J_{AA}}{J_{BB}} \right)= \xi_{1} \text{ }\xi_{2} &=& \left[ a_{1}\exp\left(-a_{2} \left|\dfrac{J_{AB}}{J_{BB}}\right|\right) \right] \left[ a_{3}-a_{4}\left(\dfrac{J_{AA}}{J_{BB}}\right)^{2}\right] \\
\label{eq_AAB_ticct}
\Delta T_{AAB} \left( \left|\dfrac{J_{AB}}{J_{BB}}\right|,\dfrac{J_{AA}}{J_{BB}} \right) = \eta_{1} \text{ }\eta_{2} &=& \left[-a_{5}\left| \dfrac{J_{AB}}{J_{BB}}\right| + a_{6}\right] \left[-a_{7}\left( \dfrac{J_{AA}}{J_{BB}}\right) + a_{8}\right]
\end{eqnarray}
Here, $\xi_{1}\equiv\xi_{1} \left( \left|\dfrac{J_{AB}}{J_{BB}}\right|,\dfrac{J_{AA}}{J_{BB}} \right)$; $\xi_{2}\equiv\xi_{2} \left( \left|\dfrac{J_{AB}}{J_{BB}}\right|,\dfrac{J_{AA}}{J_{BB}} \right)$ and \\
$\eta_{1}\equiv \eta_{1} \left( \left|\dfrac{J_{AB}}{J_{BB}}\right|,\dfrac{J_{AA}}{J_{BB}} \right)$; $\eta_{2}\equiv \eta_{2} \left( \left|\dfrac{J_{AB}}{J_{BB}}\right|,\dfrac{J_{AA}}{J_{BB}} \right)$.\\

\begin{figure}[!htb]
	\begin{center}
		\begin{tabular}{c}			
			\resizebox{10cm}{!}{\includegraphics[angle=0]{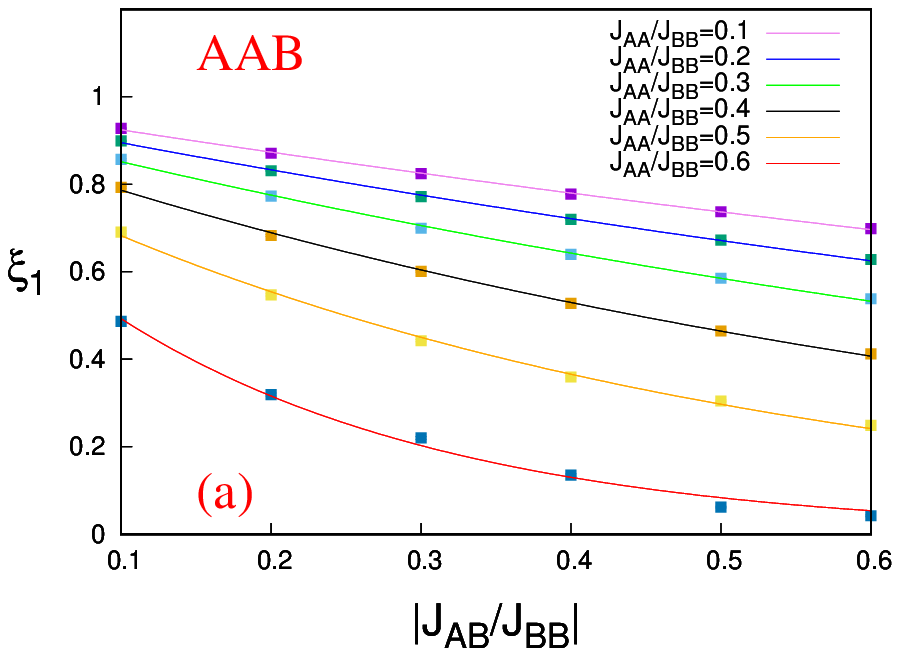}}
			\resizebox{10cm}{!}{\includegraphics[angle=0]{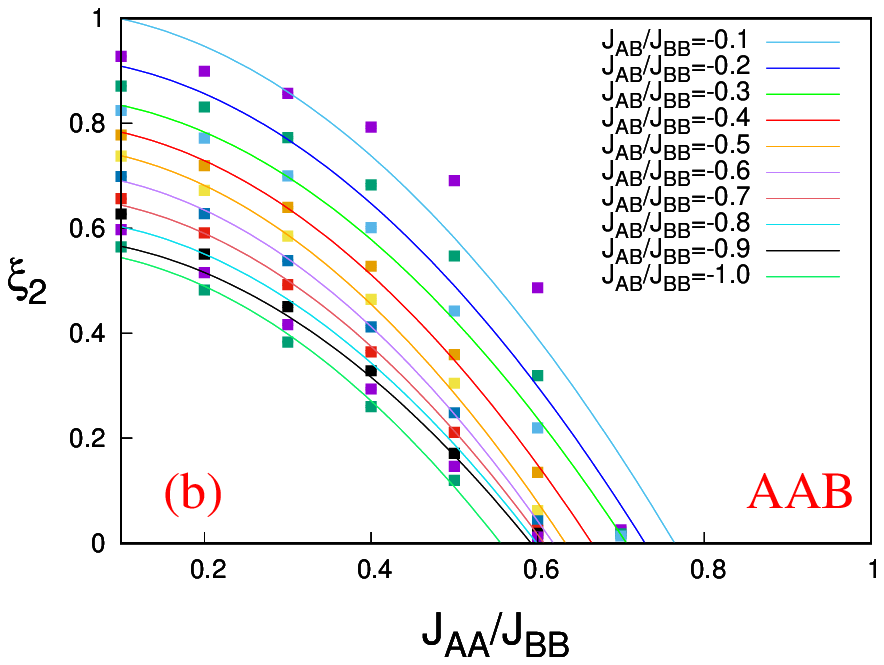}}\\
			\resizebox{10cm}{!}{\includegraphics[angle=0]{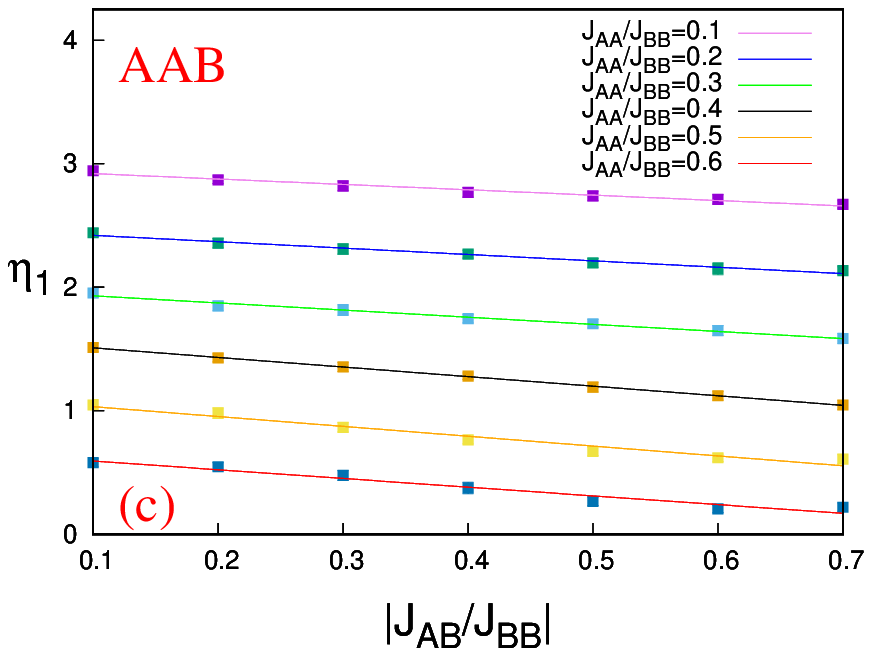}}
			\resizebox{10cm}{!}{\includegraphics[angle=0]{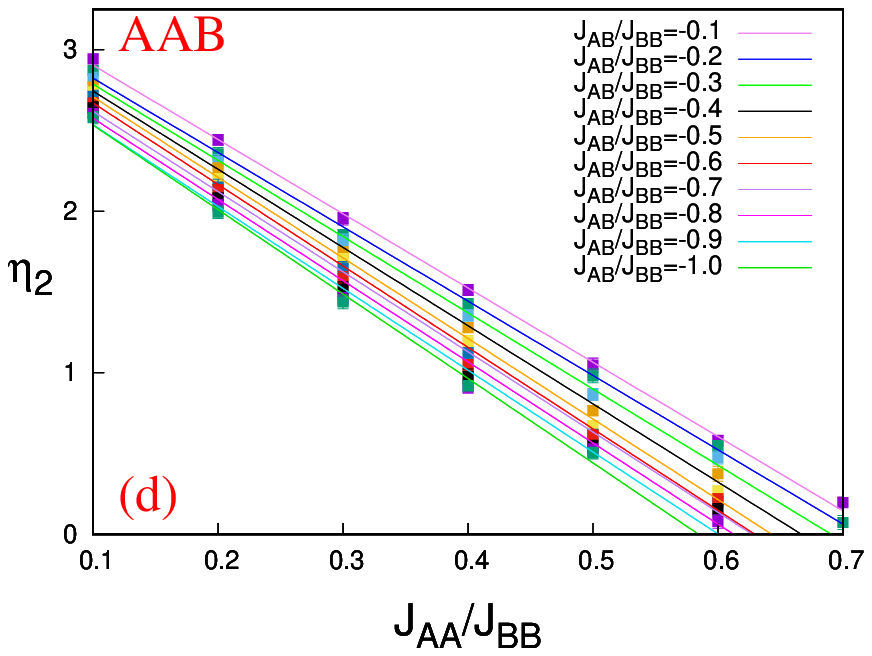}}\\
		\end{tabular}
		\caption{ (Colour Online) Functional dependence of different functions in the proposed functional forms of IARRM [(a) and (b)] and TICCT [(c) and (d)] for AAB type system [Refer to Equations \ref{eq_AAB_iarrm} and \ref{eq_AAB_ticct}, respectively]. Where the errorbars are not visible, they are smaller than the point markers.}		
		\label{fig_aab_iarrm_ticct}
	\end{center}
\end{figure}

Without any loss of generality, the parameters of the Equation \ref{eq_AAB_iarrm}, can be taken to be: $a_{1} \equiv f_{1}(J_{AA}/J_{BB})$ ; $a_{2} \equiv f_{2}(J_{AA}/J_{BB})$ ; $a_{3} \equiv f_{3}(\left|J_{AB}/J_{BB}\right|)$ and $a_{4} \equiv f_{4}(\left|J_{AB}/J_{BB}\right|)$ . Similarly for the Equation \ref{eq_AAB_ticct}, the parameters may be considered as: $a_{5} \equiv f_{5}(J_{AA}/J_{BB})$ ; $a_{6} \equiv f_{6}(J_{AA}/J_{BB})$ ; $a_{7} \equiv f_{7}(\left|J_{AB}/J_{BB}\right|)$ ; $a_{8} \equiv f_{8}(\left|J_{AB}/J_{BB}\right|) $ . A Graphical representation is in order. For the AAB type configuration, the dependence of $\xi_{1}$, $\xi_{2}$, $\eta_{1}$ and $\eta_{2}$ on respective controlling parameters can be found out in Figure \ref{fig_aab_iarrm_ticct}.

For the ABA type trilayered system, the equations for IARRM, $\mu_{ABA}$ and TICCT, $\Delta T_{ABA}$ can be written as:
\begin{eqnarray}
\label{eq_ABA_iarrm}
\mu_{ABA} \left( \left|\dfrac{J_{AB}}{J_{BB}}\right|,\dfrac{J_{AA}}{J_{BB}} \right)= \xi_{3} \text{ }\xi_{4} &=& \left[ p_{1}\exp\left(-p_{2} \left(\dfrac{J_{AB}}{J_{BB}}\right)^{2}\right) \right] \left[ p_{3}-p_{4}\left(\dfrac{J_{AA}}{J_{BB}}\right)^{2}\right] \\
\label{eq_ABA_ticct}
\Delta T_{ABA} \left( \left|\dfrac{J_{AB}}{J_{BB}}\right|,\dfrac{J_{AA}}{J_{BB}} \right) = \eta_{3} \text{ }\eta_{4} &=& \left[-p_{5}\left| \dfrac{J_{AB}}{J_{BB}}\right| + p_{6}\right] \left[-p_{7}\left( \dfrac{J_{AA}}{J_{BB}}\right) + p_{8}\right]
\end{eqnarray}
Here, $\xi_{3}\equiv\xi_{3} \left( \left|\dfrac{J_{AB}}{J_{BB}}\right|,\dfrac{J_{AA}}{J_{BB}} \right)$; $\xi_{4}\equiv\xi_{4} \left( \left|\dfrac{J_{AB}}{J_{BB}}\right|,\dfrac{J_{AA}}{J_{BB}} \right)$ and \\
$\eta_{3}\equiv \eta_{3} \left( \left|\dfrac{J_{AB}}{J_{BB}}\right|,\dfrac{J_{AA}}{J_{BB}} \right)$; $\eta_{4}\equiv \eta_{4} \left( \left|\dfrac{J_{AB}}{J_{BB}}\right|,\dfrac{J_{AA}}{J_{BB}} \right)$.

\begin{figure}[!htb]
	\begin{center}
		\begin{tabular}{c}			
			\resizebox{10cm}{!}{\includegraphics[angle=0]{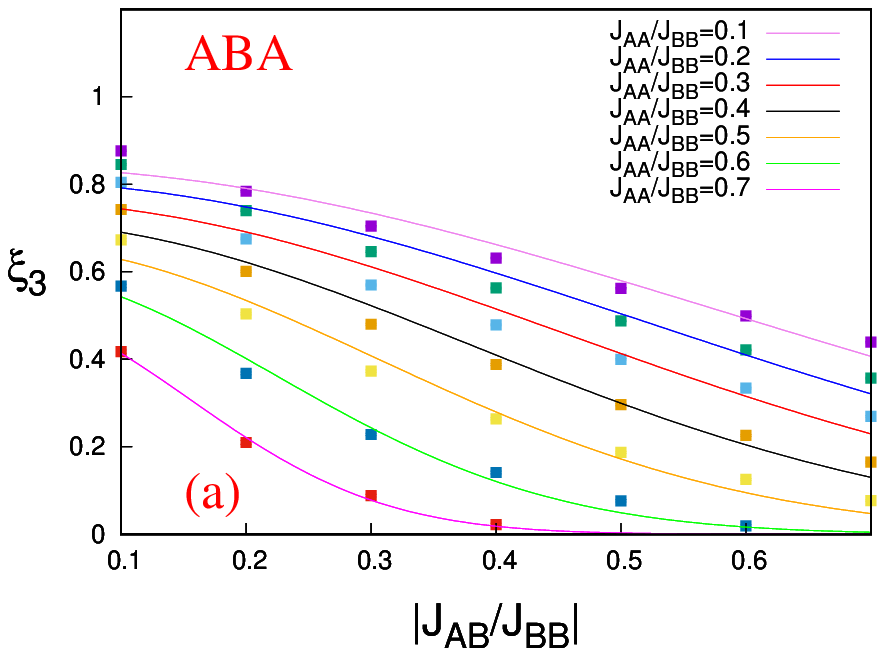}}
			\resizebox{10cm}{!}{\includegraphics[angle=0]{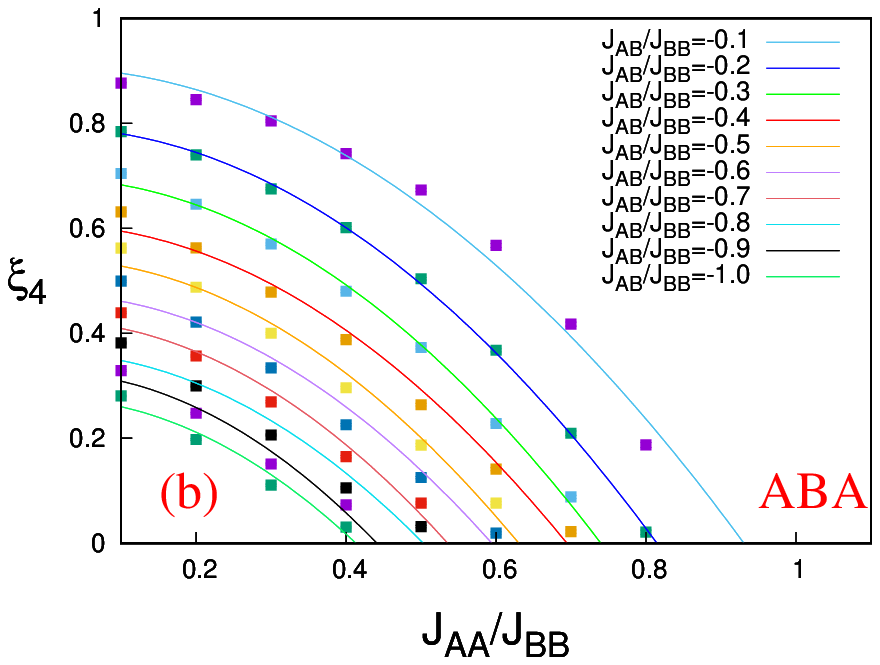}}\\
			\resizebox{10cm}{!}{\includegraphics[angle=0]{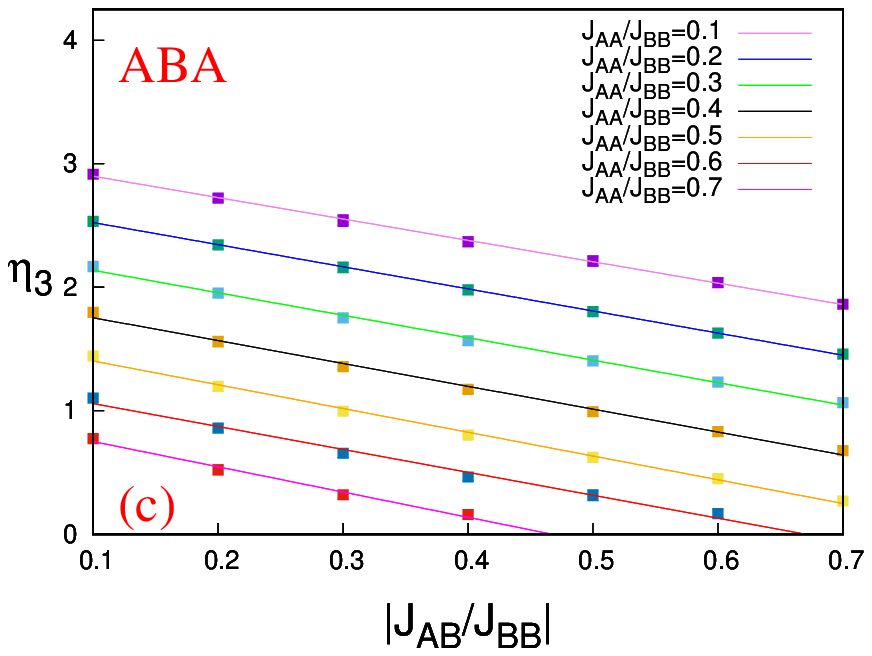}}
			\resizebox{10cm}{!}{\includegraphics[angle=0]{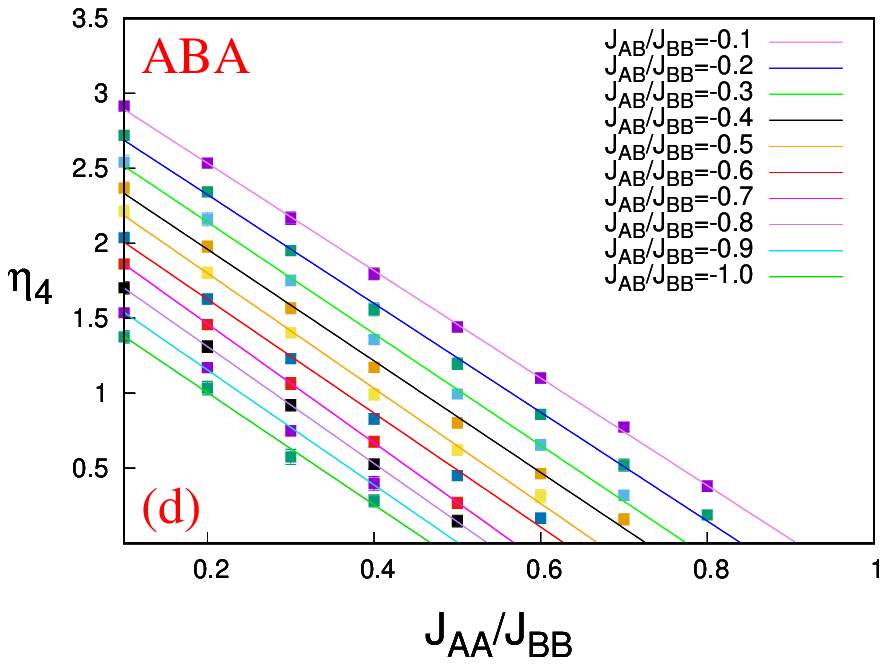}}
		\end{tabular}
		\caption{ (Colour Online) Functional dependence of different functions in the proposed functional forms of IARRM [(a) and (b)] and TICCT [(c) and (d)] for ABA type system [Refer to Equations \ref{eq_ABA_iarrm} and \ref{eq_ABA_ticct}, respectively]. Where the errorbars are not visible, they are smaller than the point markers.}		
		\label{fig_aba_iarrm_ticct}
	\end{center}
\end{figure}

Extending the earlier argument, the parameters of the Equation \ref{eq_ABA_iarrm} for IARRM in ABA trilayer, can be taken as: $ p_{1} \equiv g_{1}(J_{AA}/J_{BB}) $ ; $ p_{2} \equiv g_{2}(J_{AA}/J_{BB}) $ ; $ p_{3} \equiv g_{3}(|J_{AB}/J_{BB}|) $ ; $ p_{4} \equiv g_{4}(|J_{AB}/J_{BB}|) $ and the parameters af the Equation \ref{eq_ABA_ticct} as: $ p_{5} \equiv g_{5}(J_{AA}/J_{BB}) $ ; $ p_{6} \equiv g_{6}(J_{AA}/J_{BB}) $ ; $ p_{7} \equiv g_{7}(|J_{AB}/J_{BB}|) $ ; $ p_{8} \equiv g_{8}(|J_{AB}/J_{BB}|) $ . Following the earlier approach, for the AAB type configuration, the functional dependence of $\xi_{3}$, $\xi_{4}$, $\eta_{3}$ and $\eta_{4}$ on respective controlling parameters can be found out in Figure \ref{fig_aba_iarrm_ticct}.

The estimates of error while fitting in all the above equations are obtained through a combination of, asymptotic errors in fitting and errors obtained via Jackknife method while fitting.\\
\vspace{10pt}
\begin{center}{\Large \textbf {E. Phase diagram:}}\end{center}
\vspace{10pt}

\begin{figure}[!htb]
	\begin{center}
		\begin{tabular}{c}
			\resizebox{10cm}{!}{\includegraphics[angle=0]{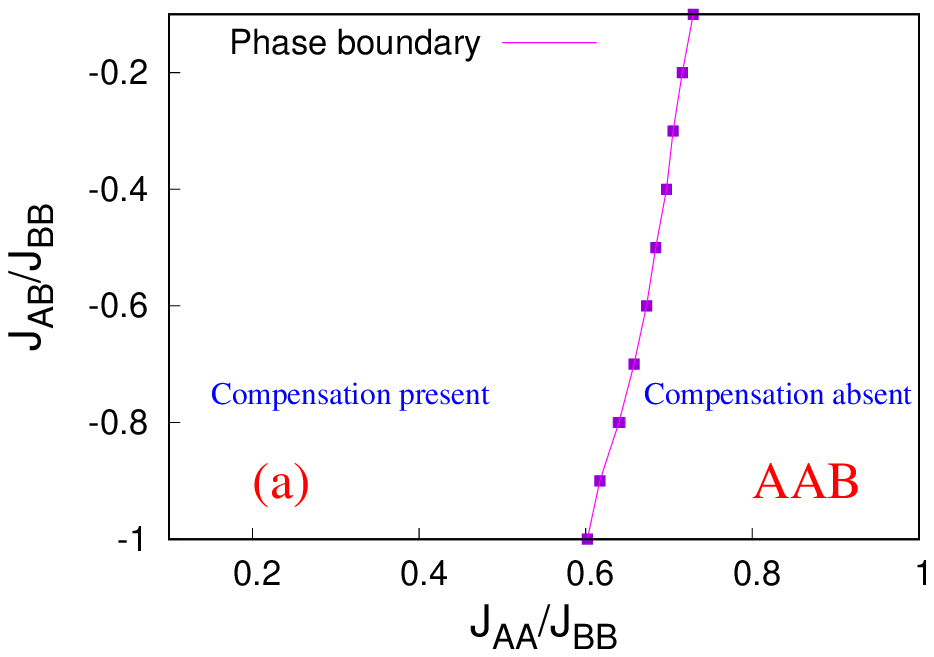}}
			\resizebox{10cm}{!}{\includegraphics[angle=0]{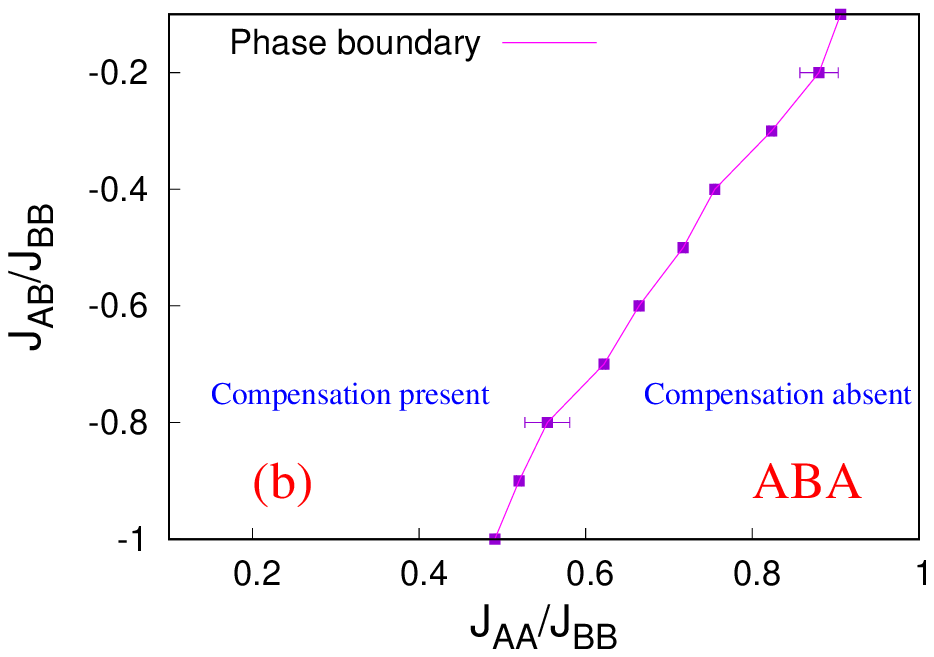}}
		\end{tabular}
		\caption{ (Colour Online) Phase diagrams for (a) AAB (b) ABA configurations. The squares were obtained through conventional MC simulations. In both the cases, the lines mark the separation between a ferrimagnetic phase with compensation (to the left) and a ferrimagnetic phase without compensation (to the right). Where the errorbars are not visible, they are smaller than the point markers.}		
		\label{fig_phasecurve}
	\end{center}
\end{figure}

Repetition of the procedure of Figure \ref{fig_zeromag_bifurcation}, for other values of $J_{AB}/J_{BB}$, one can obtain a phase diagram in the Hamiltonian paramter space. The phase seperation curve divides the parameter space into two distinct regions of interest, for both AAB and ABA configurations. One is a ferrimagnetic phase for which there is no compensation at any temperature and the other is a ferrimagnetic phase with a compensation point at a certain temperature, $T_{comp}$. These results are presented in Figure \ref{fig_phasecurve}(a) for the AAB trilayer and in Figure \ref{fig_phasecurve}(b) for the ABA trilayer. In both diagrams, to the left of the phase boundary exists a ferrimagnetic phase with compensation and to the right exists a ferrimagnetic phase without compensation.

We see that the compensation is only possible if $J_{AA}<J_{BB}$ and in both configurations, Compensation is always observed for smaller $J_{AA}/J_{BB}$, irrespective of the value of $J_{AB}/J_{BB}$. But the range of values of $J_{AA}/J_{BB}$, for which compensation occurs, shrinks as $J_{AB}/J_{BB}$ increases. The main observation is that, both the critical temperature ($T_{crit}$) and the compensation temperature ($T_{comp}$) decreases with decrease in the strength of either of the ferromagnetic or the antiferromagnetic ratio or both. While increasing, after reaching a certain threshold, for both the interaction ratios, these two temperatures merge. 

\begin{figure}[!htb]
	\begin{center}
		\begin{tabular}{c}
			\resizebox{10cm}{!}{\includegraphics[angle=0]{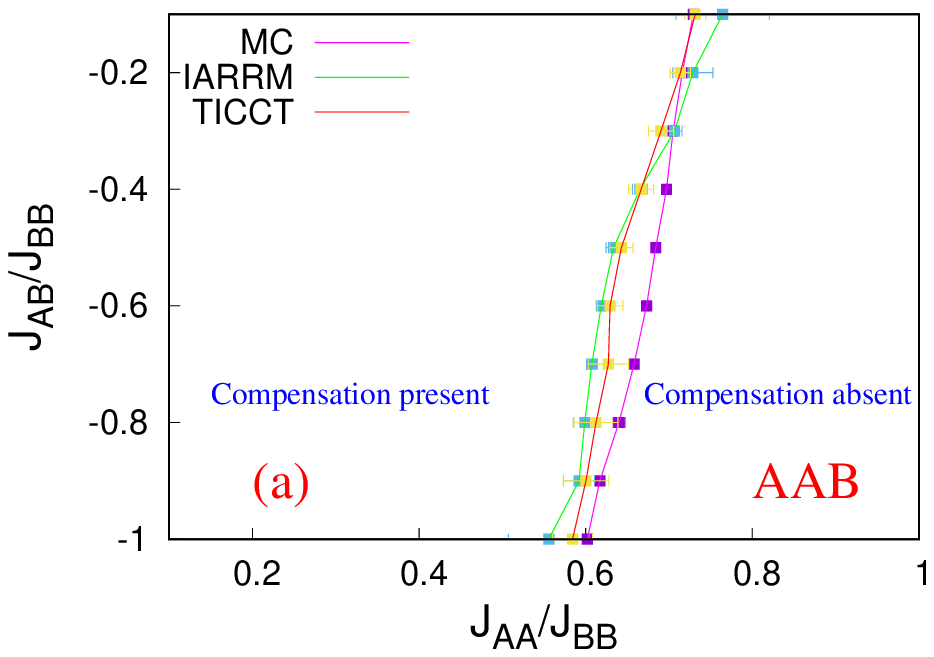}}
			\resizebox{10cm}{!}{\includegraphics[angle=0]{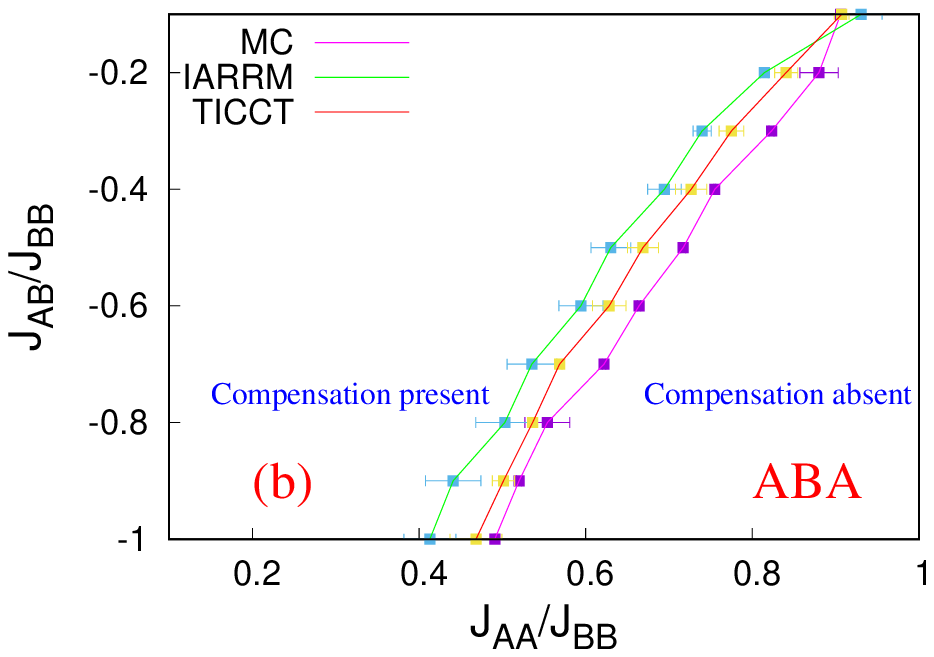}}
		\end{tabular}
		\caption{ (Colour Online) Combined phase diagrams for (a) AAB (b) ABA configurations. These plots show a good agreement between MC simulations and empirical formulation by IARRM and TICCT. Where the errorbars are not visible, they are smaller than the point markers.}		
		\label{fig_phasecurve_combined}
	\end{center}
\end{figure} 

The zeroes of the functions in Equation \ref{eq_AAB_iarrm} and Equation \ref{eq_AAB_ticct} constitute the phase seperation curves in Figure \ref{fig_phasecurve_combined}(a) for the empirical formulation of AAB configuration by IARRM and TICCT respectively. The same, for the ABA variant, can be found out in Figure \ref{fig_phasecurve_combined}(b). We can see that the alternative description by IARRM and TICCT are in good agreement with the results obtained by MC simulation. 
\vspace{10pt}
\begin{center}{\Large \textbf {V. Conclusion}}\end{center}
\vspace{10pt}

For odd number of layers in layered ferrimagnets, neither cite dilution \cite{Santos} nor mixed-spin cases \cite{Diaz1} are necessary for observing compensation on square trilayers. So the type of systems in this article, although with sublayers on triangular lattice, is among the simplest layered ferrimagnetic systems for compensation. The single spin-flip algorithm takes fluctuations into account, unlike MFA and also produces accurate results. \\

After going through the magnetic responses of sublayers in Section A and morphological studies in Section B, we are now in a position to qualitatively discuss about why compensation should be expected in certain cases for the trilayered Ising AAB and ABA type configurations with triangular sublayers. In all the cases, the magnetization for the layer with dominant in-plane coupling strength (with B-atoms) saturates rapidly below the critical point, compared to the other two layers. In Figure \ref{fig_mag_response} (a): for the AAB configuration, the magnetization of the bottom B-layer saturates rapidly while decerasing the temperature below the critical point [Refer to Figure \ref{fig_aab_morpho1}]. The rate of \textit{decrease} of magnetization for both the mid and top A-layers is smaller compared to the bottom layer (the antiferromagnetic coupling strength is weaker compared to the dominant in-plane coupling strength). So the average magnetisation, initially follows the trend of B-layer and becomes positive, attains a maxima. But, as at further lower temeperatures the magnetizations of mid and top layers gain significant magnitude, cumulatively the three sublayered-magnetizations cancel each other out, leading to the compensation point with zero (average) magnetization. Further lowering of the temperature leads to the saturation of magnetization of the A-layers. The average magnetization, similarly saturates to $-1/3$ . In Figure \ref{fig_mag_response} (b): for the AAB configuration, the magnetization for the bottom B-layer and mid A-layer saturates oppositely at a competitive rate (the magnitude of antiferromagnetic coupling strength is equal to the dominant in-plane ferromagnetic coupling). That is why, the average magnetisation for a few temeperature points below the critical temperature, still stays at zero [Refer to Figure \ref{fig_aab_morpho2} and \ref{fig_aab_morpho3}]. When the magnetization of top A-layer starts gaining significant magnitude, the average magnetisation follows the similar trend while saturating to $-1/3$. In Figure \ref{fig_mag_response} (c): for the ABA configuration, the reason of such magnetic response is quite similar to Figure \ref{fig_mag_response} (a). The magnetization of the mid B-layer saturates rapidly below the critical point. The magnetization of the top and bottom A-layers saturate at a slower rate below the critical point (the magnetization curves completely overlap) [Refer to Figure \ref{fig_aba_morpho1}]. Here also the antiferromagnetic coupling strength is weaker compared to the dominant in-plane coupling strength. Initially, the average magnetization follows the trend of the mid-layer, attains a maxima and then, as the surface layers obtain significant magnitude, the sublayered magnetizations cancel out each other and we have compensation point with zero magnetization. The average magnetization, thereafter follows the trend of surface layers and saturates to $-1/3$ . In Figure \ref{fig_mag_response} (d): for the ABA configuration, the magnetization for the mid B-layer and surface A-layers saturates oppositely [Refer to Figure \ref{fig_aba_morpho2} and \ref{fig_aba_morpho3}]. Here the magnitude of antiferromagnetic coupling strength is equal to the dominant in-plane ferromagnetic coupling. The average magnetisation monotonically decreases following the trend of surface layers while saturating to $-1/3$.\\

The magnetic responses in Figure \ref{fig_mag_response_vsize} conclude, the compensation temperature is independent of system size, after reaching a certain threshold. So we should not expect a scaling bahaviour for compensation temperatures. But around critical point, we have to take care of finite size effects. The cumulant crossing technique, used for determining critical temperatures, is phenomenological. It saves us from additional non-linear fitting procedures while doesn't compromise much on accuracy \cite{Ferrenberg}. Finally we see the traditional phase diagram in Figure \ref{fig_phasecurve} from Monte Carlo simulational data in Hamiltonian parameter space ($J_{AB}/J_{BB}$ vs. $J_{AA}/J_{BB}$). For the AAB configuration, we see that the phase boundary is much less inclined to the $J_{AB}/J_{BB}$ axis than the ABA configuration. The ratio of ferromagnetic to antiferromagnetic bonds per site is 7:1 for the mid A-layer and 6:1 for the bottom B-layer in AAB configuration compared to 3:1 in the mid B-layer and 6:1 in the bottom A-layer in ABA configuration. While in AAB, top A-layer has no antiferromagnetic bond, per site against a bond-ratio of 6:1 per site, in the top A-layer of ABA system. Greater number of antiferromagnetic bonds is responsible for the proneness to change of phase boundary with change in the values of $J_{AB}/J_{BB}$ in ABA configuration. A visual inspection of the phase boundaries for the AAB and ABA configrations of trilayered square Ising ferrimagnetic system in \cite{Diaz3} reveals, these curves are much more inclined to $J_{AB}/J_{BB}$ axis compared to their triangular counterpart, in the same parameter space and the reason is same: relative increase of the antiferromagnetic bonds compared to ferromagnetic ones in square lattice than triangular lattice. The following Maxwell's relation $\left( \dfrac{\partial S}{\partial H}\right)_{T}=\left( \dfrac{\partial M}{\partial T}\right)_{H}$ relates Magnetic entropy change, $\Delta S$ and change in Magnetization, $\Delta M$, with H and T being the applied magnetic field and the temperature of the system, respectively. So, for an abrupt change in magnetisation around the compensation point, large MCE may be expected.

The lowering of compensation temperature is particularly useful in MCE. In MCE, low temperatures $\sim \mu$K has already been achieved. If ferrimagnetic materials with compensation temperatures (similar to the ones of the current article) be used in MCE instead of traditional materials, with the same number of magnetization-demagnetization steps, we may achieve even lower temperature. Such layered magnetic materials with compensation phenomenon are economically cheaper compared to rare-earths, thus making them suitable candidates for MCM. Particularly useful is the description in terms of empirical formulations (IARRM and TICCT) of bulk properties. The agreement of MC simulations and empirical descriptions is acceptable within the demarcated region in phase diagram where technological applications involving presence of compensation would take place [Figure \ref{fig_phasecurve_combined}]. The proposed formulae of IARRM and TICCT, alongwith the functional forms of the parameters, can readily provide insights for range of controlling Hamiltonian parameters, without performing the entire range extensive MC simulation. This is extremely beneficial for experimentalists in designing materials for targeted purposes with required properties.
\vspace{30pt}
\begin{center} {\Large \textbf {Acknowledgements}}\end{center}
The author acknowledges financial help from University Grants Commission, India in the form of research fellowship and is grateful to Tamaghna Maitra for discussions, feedback and technical assistance. Several insightful comments and suggestions made by the anonymous referees are also acknowledged.
\vspace{50pt}
\begin{center} {\Large \textbf {References}} \end{center}

\end{document}